\documentclass[sigconf]{acmart}
\pdfoutput=1
\usepackage{hyperref}
\usepackage{enumitem}
\usepackage{array}
\usepackage{graphicx}
\usepackage{balance}
\usepackage{subfigure}
\usepackage{multirow}
\usepackage{float}
\usepackage{soul}
\usepackage{mdframed}
\usepackage{amsopn}
\usepackage{mathrsfs}
\usepackage{mathtools}
\usepackage{amsmath}
\usepackage{bbm}
\usepackage{arydshln}
\begin{document}

\title[Fairness-Aware Explainable Recommendation over Knowledge Graphs]{Fairness-Aware Explainable Recommendation\\over Knowledge Graphs}

\author[Fu et al.]{Zuohui Fu$^{\dagger*}$, Yikun Xian$^{\dagger*}$, Ruoyuan Gao${^\dagger}$, Jieyu Zhao${^\ddagger}$, Qiaoying Huang${^\dagger}$, Yingqiang Ge${^\dagger}$, Shuyuan Xu${^\dagger}$, Shijie Geng${^\dagger}$, Chirag Shah$^\S$, Yongfeng Zhang${^\dagger}$, Gerard de Melo${^\dagger}$}

\renewcommand{\authors}{Zuohui Fu, Yikun Xian, Ruoyuan Gao, Jieyu Zhao, Qiaoying Huang, Yingqiang Ge, Shuyuan Xu, Shijie Geng, Chirag Shah, Yongfeng Zhang, Gerard de Melo}

\thanks{${}^*$Both authors contributed equally to this work.}

\affiliation{
  \institution{$^{\dagger}$Rutgers University\qquad $^{\ddagger}$University of California, Los Angeles\qquad $^\S$University of Washington}
}

\email{zuohui.fu@rutgers.edu, siriusxyk@gmail.com, ruoyuan.gao@rutgers.edu, jyzhao@cs.ucla.edu}
\email{{qh55,yingqiang.ge, shuyuan.xu, sg1309}@rutgers.edu, chirags@uw.edu, yongfeng.zhang@rutgers.edu, gdm@demelo.org}

\begin{abstract}
There has been growing attention on fairness considerations recently, especially in the context of intelligent decision making systems. Explainable recommendation systems, in particular, may suffer from both explanation bias and performance disparity. 
In this paper, we analyze different groups of users according to their level of activity, and find that bias exists in recommendation performance between different groups. 
We show that inactive users may be more susceptible to receiving unsatisfactory recommendations, due to insufficient training data for the inactive users, and that their recommendations may be biased by the training records of more active users, due to the nature of collaborative filtering, which
leads to an unfair treatment by the system. 
We propose a fairness constrained approach via heuristic re-ranking to mitigate this unfairness problem in the context of explainable recommendation over knowledge graphs. We experiment on several real-world datasets with state-of-the-art knowledge graph-based explainable recommendation algorithms. The promising results show that our algorithm is not only able to provide high-quality explainable recommendations, but also reduces the recommendation unfairness in several respects.
\end{abstract}
\copyrightyear{2020}
\acmYear{2020}
\setcopyright{acmcopyright}
\acmConference[SIGIR '20]{Proceedings of the 43rd International ACM SIGIR Conference on Research and Development in Information Retrieval}{July 25--30, 2020}{Virtual Event, China}
\acmBooktitle{Proceedings of the 43rd International ACM SIGIR Conference on Research and Development in Information Retrieval (SIGIR '20), July 25--30, 2020, Virtual Event, China}
\acmPrice{15.00}
\acmDOI{10.1145/3397271.3401051}
\acmISBN{978-1-4503-8016-4/20/07}
\settopmatter{printacmref=true}
\maketitle

\section{Introduction}

Compared with traditional recommendation systems (RS), explainable recommendation is capable of not only providing high-quality recommendation results but also offering personalized and intuitive explanations \cite{zhang2018explainable}, which are important for e-commerce and social media platforms. 
However, current explainable recommendation models leave two major concerns in terms of fairness. First, the model discriminates unfairly among the users in terms of recommendation performance. And 
second, the model may further discriminate between users in terms of explanation diversity.
In this paper, we consider the fairness issues of both performance imbalance and explanation diversity in explainable recommendation, which arises from the fact that there may be groups of users who
are less noticeable on a platform, e.g., due to inactivity, making them less visible to the learning algorithms. 

One reason for this relates to the issue of data imbalance. Some users are disinclined to make a large number of purchases, which leads to insufficient historical user--item interactions.
For instance, on e-commerce platforms such as Amazon, eBay, or Taobao, economically disadvantaged groups often make fewer purchases in light of their limited income and credit opportunities \cite{ge2020learning}.
Under such circumstances, when making recommendation decisions, explainable RS models will be subject to algorithmic bias. The lack of user--item interactions implies that the corresponding user preferences are barely captured, causing weak visibility of such users to the RS model.
This leads to the risk of such users being treated unfairly in terms of both recommendation performance and explanation diversity.
In this paper, we aim at alleviating such algorithmic bias and improving the fairness of explainable recommendations.

Unfortunately, it is challenging to study fairness in recommendation systems due to the lack of unifying definitions and means of quantifying unfairness.
\citet{Farnadi2018AFH} claim that no model can be fair in every aspect of metrics.
Previous work has explored the fairness problem in recommendation from the perspective of selection aspects \cite{Geyik2019,singh2018fairness,Schnabel2016RecommendationsAT},
marketing bias \cite{Wan2019AddressingMB}, popularity bias \cite{yang2018unbiased}, multiple stakeholders \cite{Burke2017MultisidedFF} in terms of consumers and providers, among others.
Existing research on fairness has shown that \emph{protected groups}\footnote{\citet{Chen2019FairnessUU} summarizes the protected classes defined by the US Fair
Housing Act (FHA) and Equal Credit Opportunity Act (ECOA).}, defined as the population of vulnerable individuals in terms of sensitive features such as gender, age, race, religion, etc., are easily treated in a discriminatory way. 
However, it is generally not easy to obtain access to such sensitive attributes, 
as users often prefer not to disclose such personal information.
In this study, we instead consider a directly observable property, the visibility of the user to the explainable RS model, which relates to a user's level of activity on the platform, and may directly entail subpar treatment by the recommendation engine.

We are interested in solving the fairness problem on the user side specifically for knowledge graph (KG) enhanced explainable recommender systems. Since KGs preserve structured and relational knowledge, they make it easy to trace the reason for specific recommendations. KG-based approaches have thus grown substantially in popularity in explainable recommendation.
Their explicit explanations take the form of reasoning paths, consisting of a sequence of relationships that start from a user and ultimately lead to a recommended item.
State-of-the-art KG-based explainable RS methods \cite{ai2018learning, xian2019kgrl,Wang2019KGATKG,wang2018ripplenet,zhang2016collaborative,xian2020symbol} utilize rich entity and relation information within the KG to augment the modeling of user--item interactions\footnote{We interchangeably use ``user interactions" and ``user--item interactions" in the paper.}, so as to better understand the user preferences to make satisfactory recommendation decisions, accompanied by explainable reasoning paths.
However, due to the fundamental nature of collaborative filtering, current KG-based explainable recommendation methods heavily rely on users' collective historical interactions for model learning, so the recommendations and corresponding explanations tend to be more consistent with the dominating historical user interactions. Because of this, current RS methods tend to neglect the user--item interactions of less visible, inactive users, since they are easily overwhelmed by more visible, active users.

Thus, we argue that it is critical for a RS to pay attention to inactive users as well, so that they can be served with high-quality recommendations and more diverse explanations.
The connecting paths between users and recommended items are expected to be highly relevant and match past user interactions. Thus, a learning algorithm drawing on user--item path links is likely to yield better recommendation performance 
for users who have contributed more interactions. However, the remaining portion of users that are less visible to the model may end up not enjoying the same recommendation experience. In part, this can stem from a lack of a relevant user--item interaction history to accurately reveal the user preferences.
However, even if a user is not entirely inactive, the model's training on input data dominated by the more visible users can easily lead to it being biased towards the interactions made by the most active and privileged users.

In this work, 
we capture user--item interactions at both the individual and group level in terms of user--item paths. We particularly seek to understand
1) how to verify our concerns about the unfairness problems in explainable recommender systems and quantify such unfairness;
2) how to alleviate any potential algorithmic bias so as to improve the recommendation quality, while providing diverse explanations, especially for disadvantaged users, 
3) whether our fairness-aware method is able to consider both group-level fairness and individual-level fairness, and whether it possesses generalizability to multiple KG-enhanced explainable RS methods.

Based on these, our main contributions include:
\begin{itemize}[leftmargin=*]
    \item We study four e-commerce datasets from Amazon and conduct a data-driven observation analysis to assess their data imbalance characteristics. We identify unfairness owing to the difference in historical user--item interactions, and argue that current KG-based explainable RS algorithms neglect the discrepancy of user preferences, which gives rise to unfair recommendations. Additionally, we devise the \emph{group fairness} and \emph{individual fairness} criteria with regard to recommendation performance and explanation diversity.
    \item Since there are intrinsic differences in user preferences among the users due to data imbalance, our goal is not to pursue an absolute parity of recommendations and explanation diversity. Rather, we propose a fairness-aware algorithm so as to provide fair explainable diversity leading to potential items of interest for recommendations. Specifically, we formalize this as a 0--1 integer programming problem and invoke modern heuristic solving techniques to obtain feasible solutions. 
    \item 
    Our algorithm is expected to improve the recommendation quality while narrowing the disparity between different groups of users. Through extensive experiments and case studies, the quantitative results suggest that our fairness-aware algorithm provides significant improvements in both recommendation and fairness evaluation, at both the group level and individual level.
\end{itemize}
\vspace{-8pt}
\section{Related Work}
\noindent\textbf{Fairness in Decision Making.}\quad
Growing interest in fairness has arisen in several research domains. Most notably, for data-driven decision-making algorithms, there are concerns about biases in data and models affecting minority groups and individuals \cite{corbett2017algorithmic}. Group fairness, also known as demographic parity, requires that the protected groups be treated equally to advantaged groups or the general population \cite{hardt2016equality, pedreshi2008discrimination, singh2018fairness}.
In contrast, individual fairness requires that similar individuals with similar attributes be treated similarly \cite{dwork2012fairness, Biega2018EquityOA, kusner2017counterfactual, lahoti2018ifair}. Several prior works have sought to quantify unfairness both at the group and individual level \cite{Kleinberg2016InherentTI}. 
Model bias has in fact been shown to amplify biases in the original data \cite{barocas2016big,feldman2015certifying,Zhao2017MenAL}. For each specific domain, there is a need to design suitable metrics to quantify fairness and develop new debiasing methods to mitigate inequity for both groups and individuals.

\vspace{5pt}
\noindent\textbf{Fairness-aware Ranking and Recommendation.}\quad
In the field of recommendation systems, the concept of fairness has been extended to multiple stakeholders \cite{Burke2017MultisidedFF}. \citet{lin2017fairness} defined fairness measures in recommendation and proposed a Pareto optimization framework for fair recommendation. \citet{Mehrotra2018} addresses the supplier fairness in two-sided marketplace platforms and proposed heuristic strategies to jointly optimize fairness and relevance.
Different aspects of fairness have been explored.
\citet{Beutel19} investigated pairwise recommendation with fairness constraints. \citet{celis2019controlling} addressed the polarization in personalized recommendations, formalized as a multi-armed bandit problem.
As for the fairness ranking, \citet{zehlike2017fa} proposed a fair top-$k$ ranking task that ensures that the proportion of protected groups in the top-$k$ list remains above a given threshold.
\citet{singh2018fairness} presented a conceptual and computational framework for fairness ranking that maximizes the utility for the user while satisfying specific fairness constraints. \citet{Geyik2019} developed a fairness-aware ranking framework that improves the fairness for individuals without affecting business metrics.
\citet{wu2018discrimination} draw on causal graphs to detect and remove both direct and indirect rank bias, and show that a casual graph approach outperforms statistical parity-based approaches in terms of the identification and mitigation of rank discrimination. 
In our work, we are particular interested in the disparity of user visibility to modern ranking algorithms in recommendation systems. 

\vspace{5pt}
\noindent\textbf{Explainable Recommendation with Knowledge Graphs.}\quad
Explainable recommendation \cite{zhang2018explainable} has been an important direction in recommender system research. Past work has considered explaining latent factor models \cite{Zhang:2014ecom}, explainable deep models \cite{gao2019explainable}, social explainable recommendations \cite{ren2017social}, visual explanations \cite{chen2019personalized}, sequential explanations \cite{chen2018sequential}, and dynamic explanations \cite{chen2019dynamic}.
An important line of research leverages entities, relationships, and paths in knowledge graphs to make explainable decisions. Within this field, \citet{ai2018learning} incorporated TransE-based knowledge graph representations for explainable recommendation. 
\citet{Wang2019KGATKG} proposed an attention-based knowledge-aware model to infer user preferences over KGs for recommendation. \citet{xian2019kgrl} adopted reinforcement learning for path inference in knowledge graphs. \citet{chen2020jointly} improved the efficiency of KG-based recommendation based on non-sampling learning.
However, none of these works considered model bias, which may lead to both recommendations and explanations that fail to satisfy basic principles of fairness.

\section{PRELIMINARIES}
\label{sec:pre}
In this section, we introduce the relevant concepts regarding explainable recommendation over knowledge graphs.

A \textbf{knowledge graph} is defined as a set of triples with $\mathcal{G} = \{(e_\mathrm{h},r,e_\mathrm{t})\mid e_\mathrm{h},e_\mathrm{t}\in\mathcal{E},r\in\mathcal{R}\}$, where $\mathcal{E}$
is a set of entities 
and $\mathcal{R}$
is a set of relations connecting two different entities.
A relationship between a head entity $e_\mathrm{h}$ and tail entity $e_\mathrm{t}$ through relation $r$ in the graph can be represented as the triple $(e_\mathrm{h},r,e_\mathrm{t})$.
In standard recommendation scenarios, 
the subset $\mathcal{U}$ stands for the \emph{User} entities, while $\mathcal{V}$ represents \emph{item} entities ($\mathcal{U} \cap \mathcal{V} = \varnothing$). Each relation $r\in\mathcal{R}$ uniquely determines the candidate sets for its head and tail entities. For example, a ``purchase'' in e-commerce recommendation, denoted by $r_\mathrm{up}$, always has
$(e,r_\mathrm{up},e')\in\mathcal{G} \Rightarrow e\in\mathcal{U}, e'\in\mathcal{V}$.

A \textbf{pattern} $\pi$ of length $|\pi|$ in $\mathcal{G}$ is defined as the sequential composite of $|\pi|$ relations, $\pi=\{r_1 \circ r_2 \circ \cdots \circ r_{|\pi|} \mid r_i\in\mathcal{R}, i\in[|\pi|]\}$, where "$\circ$" denotes the composition operator on relations. 
A \textbf{path} with respect to a pattern $\pi$, denoted by $L^\pi$, is a sequence of entities and relations, defined as $L^\pi=\{e_0,r_1,e_1,\ldots,e_{|\pi|-1},r_{|\pi|},e_{|\pi|}\mid (e_{i-1},r_i,e_i)\in\mathcal{G}, i\in [1, |\pi|]\}$.
In the context of KG-based recommendation, we specifically consider \textbf{user--item paths} of path pattern $\pi$, denoted by $L_{uv}^{\pi}$ as a connecting path from user $u$ to item $v$, which satisfies that $e_0=u$ and $e_{|\pi|}=v$.

We also define the \textbf{user--item path distribution} over user $u$ and item set $\mathcal{V}$, denoted by $\mathcal{D}_{u,\mathcal{V}}$, to be $\mathcal{D}_{u,\mathcal{V}}(\pi) = \frac{N(\pi)}{\sum_{\pi'\in \Pi} N(\pi') }$, where $N(\pi)$ denotes the occurrence frequency of user--item paths with respect to pattern $\pi$, i.e., $|\{L_{uv}^\pi \mid v\in \mathcal{V}\}|$.
The original problem of explainable recommendation over KGs is formally defined as:
\begin{definition}{(\textbf{Explainable Recommendation over KGs})}
Given an incomplete knowledge graph $\mathcal{G}$, the goal is to recover missing facts $\{(u,r_{uv},v) \mid (u,r_{uv},v)\not\in\mathcal{G},u\in\mathcal{U},v\in\mathcal{V}\}$ such that each fact $(u,r_{uv},v)$ is associated with a user--item path $L_{uv}^{\pi}$, where item $v$ is the recommendation for user $u$ and the path $L_{uv}^{\pi}$ is the explanation for the recommendation.
\end{definition}
\begin{figure}
\centering
\includegraphics[width=1.00\linewidth]{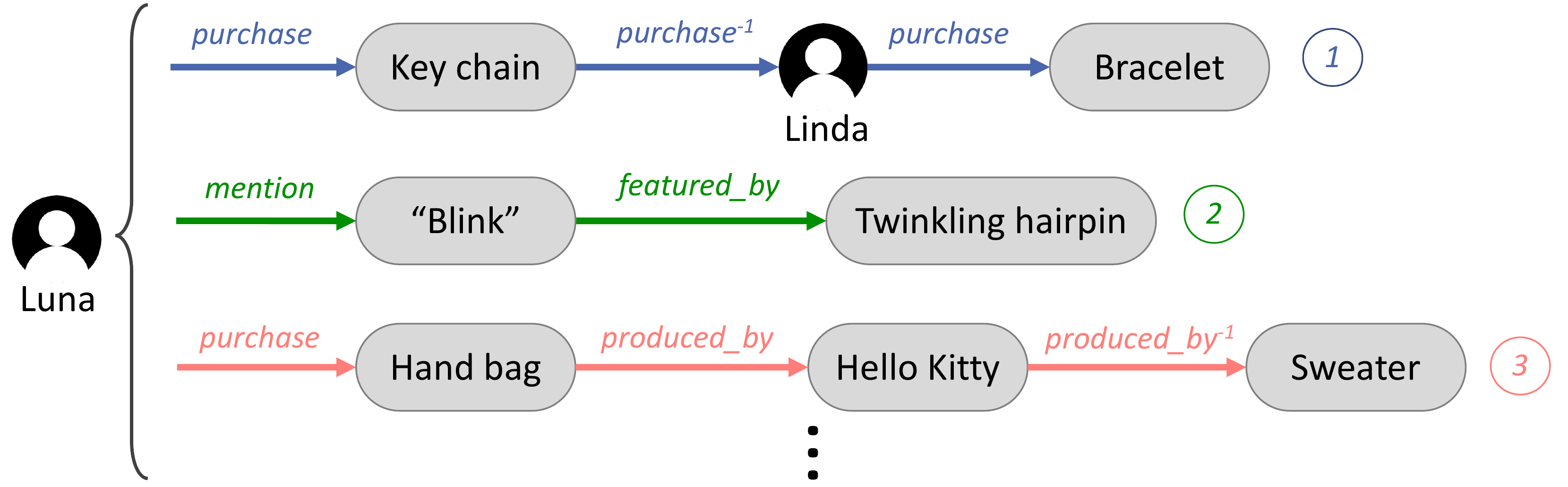}
\vspace{-17pt}
\caption{Path pattern example of user "Luna" where $-1$ means the reversed direction.}
\label{fig:path_pattern}
\vspace{-12pt}
\end{figure}
\vspace{-3pt}
\section{Motivating Fairness Concerns}\label{sec:fair_concern}

\subsection{Data Imbalance}
The traditional unfairness problem arises based on sensitive intrinsic attributes that distinguish different demographic groups  \cite{ekstrand2018all, Chen2018FairLN}. 
In this paper, we consider the visibility of users with regard to their activities in terms of user interactions. 
E.g, economically disadvantaged customers tend to make fewer purchases, leading to imbalanced data.
Current explainable models remain oblivious of such disparities in user--item interaction data.
Such imbalances, however, may lead to biased models
that exhibit unfairness with respect to
the recommendation quality and explanation diversity. 

To assess the distribution empirically, we consider Amazon datasets for 4 item categories: \emph{CDs and Vinyl}, \emph{Clothing}, \emph{Cell Phones}, and \emph{Beauty}. 
Further details of this data are given in Sec.~\ref{sec:exp}. 

Table \ref{tab:train_stats} shows the distribution of the number of items purchased in the four datasets. 
We observe that although the most active users tend to purchase more items, the majority of consumers are inactive users who are easily disregarded by commercial recommendation engines. Therefore, it is indispensable to devise techniques to better serve such users, and it can indeed also make sense economically to serve higher-quality recommendations to them with the hope of enticing them to make further purchases.

\subsection{Path Distributions as a Cause of Unfairness}
Imbalanced data can easily lead to biased models.
For explainable recommendation over KGs, the models generally consider paths along nodes in the KG 
as pertinent signals for recommendation \cite{hu2018leveraging}. 
User--item paths in the KG can directly serve as explanations that provide the reason why an item is recommended \cite{xian2019kgrl}.
For instance, in Fig.~\ref{fig:path_pattern}, in the first path pattern, Luna may wish to purchase the same bracelet as another user, since both have purchased the same key chain. Luna might also appreciate the twinkling hairpin, since its shiny feature overlaps with her review comment. Finally, Luna may consider purchasing a sweater because it matches the brand of a previous purchase, as 
in the $3$rd path.

Instead of considering particular paths along specific nodes, we can also consider just the relations involved in the paths to observe which general relational structures are serving as explanations across different paths with similar semantics. The three specific paths in Fig.~\ref{fig:path_pattern} can be viewed as instances of three different path patterns as defined in Section \ref{sec:pre}.
Thus, the distribution of user--item paths with regard to different path patterns can shed light on the diversity of explanations.

\begin{figure}[t]
\begin{minipage}[t]{1.02\linewidth}
    \includegraphics[width=0.52\linewidth]{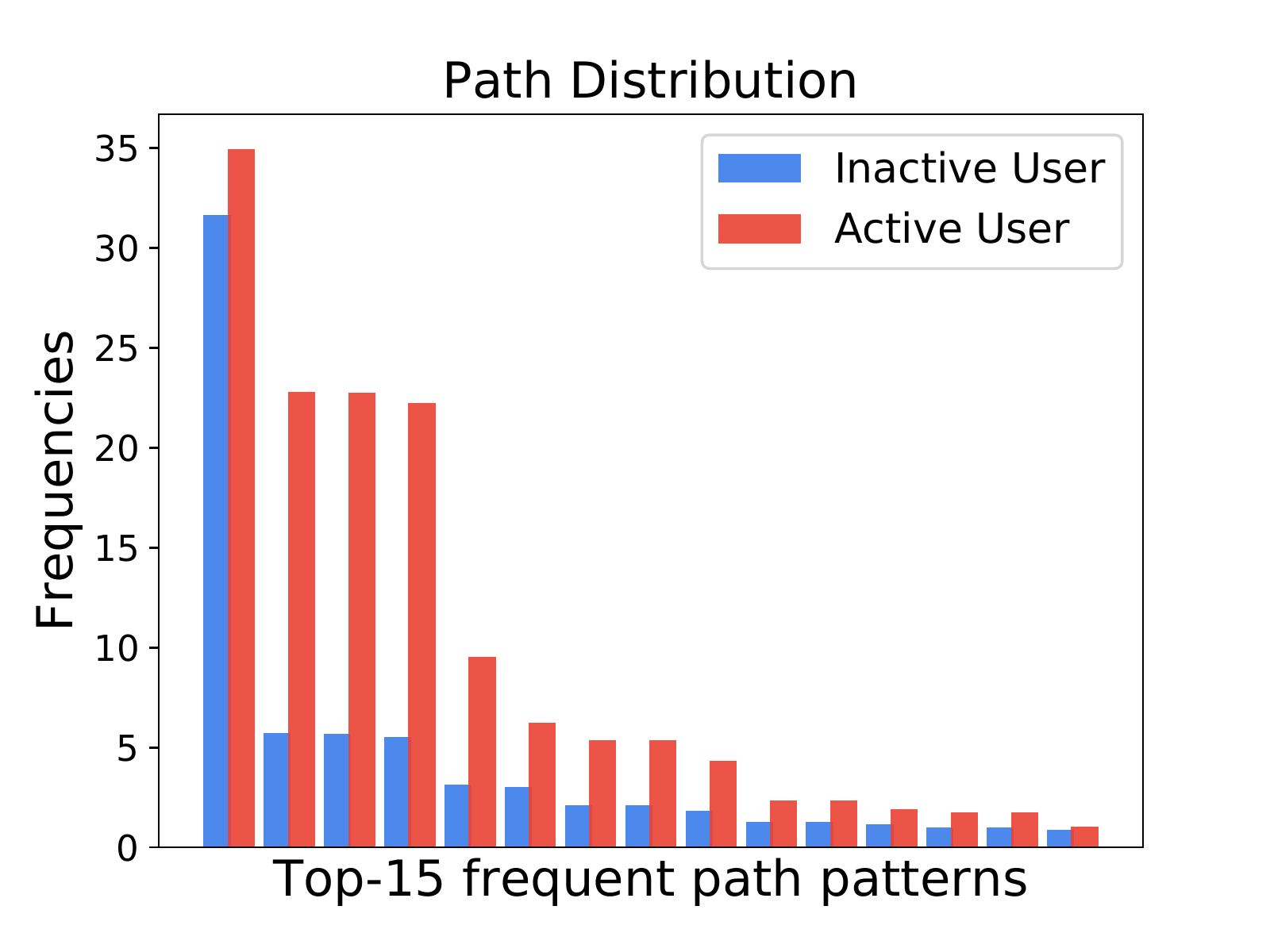}
    \includegraphics[width=0.52\linewidth]{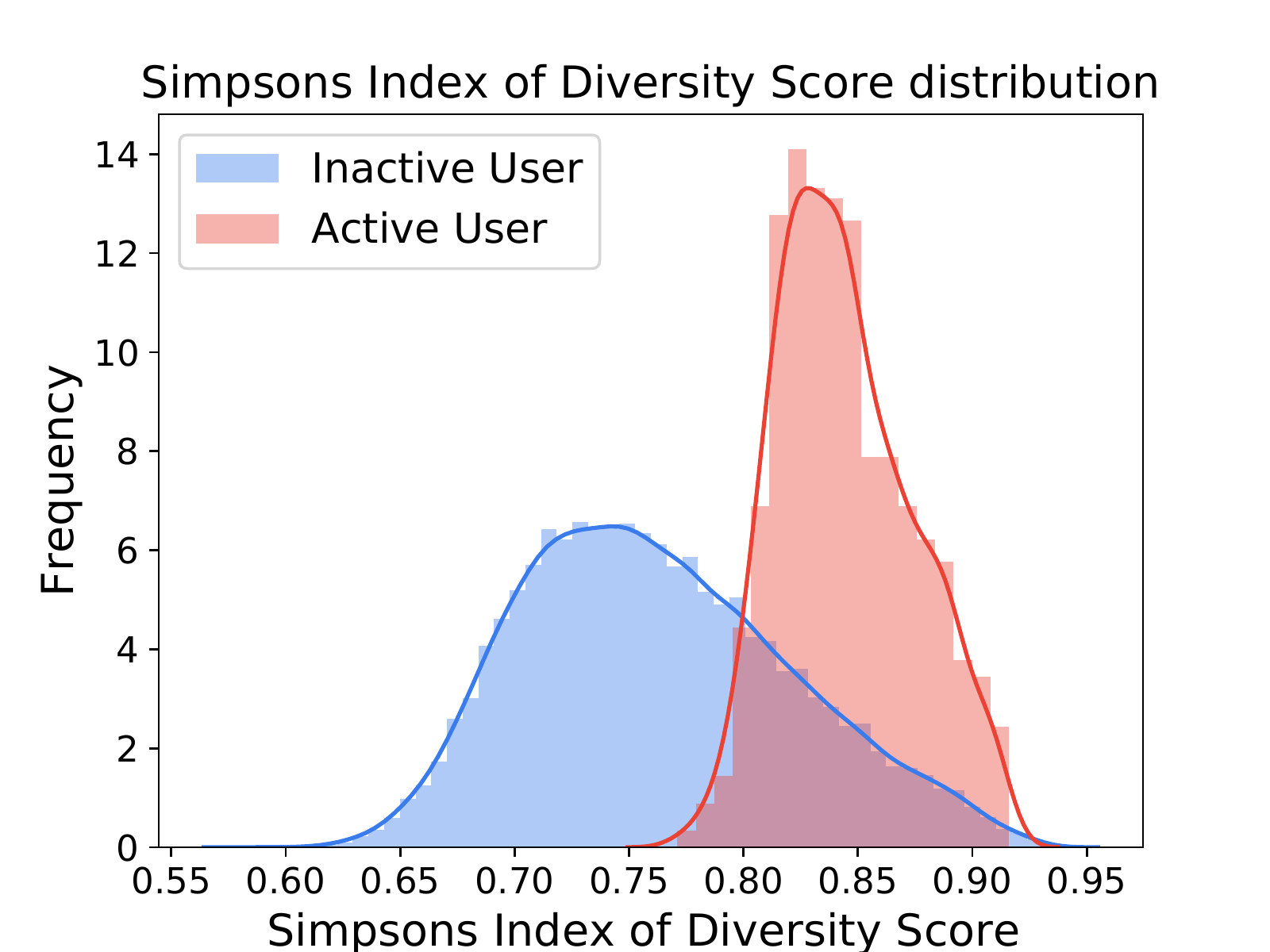}
    \label{f2}\\
     (a) Path distribution over patterns \hspace{0.3in} (b) $SID$ score distribution
    \vspace{-7pt}
    \caption{Statistics of Amazon \emph{Beauty} dataset. The same trends are observed for the other three datasets.}
    \label{fig:dataset_stats}
\end{minipage} 
\end{figure}

We claim that the divergence of user--item path distributions between two groups is an essential factor leading to unfair recommendation performance and a disparity in the diversity of explanations. 
To investigate this, we compute a series of statistics pertaining to the path distributions and recommendation quality on the aforementioned Amazon data
as follows:
\begin{enumerate}[leftmargin=*]
\item Fig.~\ref{fig:dataset_stats} (a) plots the path distribution over the top-$15$ most frequent path patterns. We consider the top 5\% of users with the largest number of purchases for each category in the training set as the \emph{active} group,
while the remaining users are considered \emph{inactive} users.
We observe divergent distributions between active and inactive users according to this group division. Although the inactive group constitutes the majority of users, their user--item path patterns lack diversity. We shall see that this can lead to unfair recommendation performance.
\item Table \ref{tab:eval}, described later in further detail, provides experimental results for a number of recommendation algorithms, with separate columns for active vs.\ inactive users. We observe that the inactive group obtains far lower scores compared to the active group, which consists of only 5\% of users. 
Thus, the performance for the vast majority of users is sacrificed.
\end{enumerate}

\noindent However, current KG-based recommendation approaches \cite{ai2018learning, Wang2019KGATKG,xian2019kgrl} neglect the distribution of paths connecting users and items. Their heuristic sampling strategy 
fits the overall path distribution, which is highly skewed.

\subsection{Quantifying Diversity}
\label{sec:diversity_simpson}
To better assess the difference in path distribution between two groups, we introduce \emph{Simpson's Index of Diversity} ($SID$) \cite{simpson1949measurement} as a metric that quantifies the unfairness. $SID$ is 
often used to quantify the biodiversity of a habitat in ecological science, taking into account the number of types (e.g., species) present, as well as the abundance of each type.
Specifically, the two main factors taken into account to measure diversity are \emph{richness} and \emph{evenness}. Richness is the number of different species present and evenness compares the similarity of the population size of each of the species present.
A habitat or community dominated by one or two species is considered to be less diverse than one in which several different species are similarly abundant. 
Note that the alternative Shannon index is more sensitive to the size of species. Conversely, Simpson's index emphasizes the dominant species more compared to the Shannon index (also called Shannon entropy). In other words, Shannon entropy is a proper manifestation mainly for species \emph{richness}, while $SID$ takes into account both \emph{richness} and \emph{evenness} measuring both absolute diversity of species and relative abundance of species. 

We conduct the corresponding analysis of our Amazon data assuming each user represents a unique community, while the patterns of corresponding user--item paths denote the species within such a community.
Formally, the $SID$ measures the probability that two randomly selected individual user--item paths belong to the same user--item path pattern. The probability of obtaining the same pattern in two random draws is defined as\footnote{If the size of dataset is very large, sampling without replacement provides approximately the same result, however, when the size of dataset is small, the difference can be substantial.}:
\begin{align}\label{eq:simpson}
   \mathcal{T}_{uv}(R, N) &= 1 - \frac{\sum_{i=1}^{R}n_{i}(n_{i} - 1)}{N(N-1)},
\end{align}
\noindent where $R$ represents the number of path patterns for the specified user $u$ to the item $v$, $n_{i}$ denotes the number of such paths belonging to the $i$-th path pattern, and $N$ is the total number of user--item paths originating from the user. The $SID$ value $\mathcal{T}_{uv}(R, N)$ ranges between 0 and 1, with larger values indicating a greater path pattern diversity. In our setup, we compute $\mathcal{T}_{uv}(R, N)$ for each user based on sampling without replacement. 
The $SID$ distribution plotted in Fig.~\ref{fig:dataset_stats} (b) shows that the inactive group of users have less diversity in their path patterns compared to the active users.
We shall later invoke Simpson's Index of Diversity in our fairness algorithm.

\begin{table}[t]
\small
\begin{tabular}{c|llll}
\hline
Dataset & {\textbf{CDs \& Vinyl}} & {\textbf{Clothing}} & {\textbf{Cell Phones}} & {\textbf{Beauty}} \\
\hline 

$ n < 4$            & 0    & 0     & 0     & 0  \\
$4 \leq n < 5$      & 18.1K & 15.3K & 12.0K & 7.1K\\
$5 \leq n < 6$      & 20.0K & 13.6K & 0.9K  & 6.9K \\
$6 \leq n < 9$      & 16.4K & 7.3K  & 4.7K  & 4.7K \\
$9 \leq n < 15$     & 11.2K     & 2.3K  & 1.1K  & 2.5K\\
$15 \leq n < 30$    & 3.5K   & 445   & 267   & 886  \\
$ n \geq 30$        & 2.1K  & 36    & 55    & 233 \\

\hline
\end{tabular}
\caption{Number of users located at different purchase thresholds(as $n$ represents) in the training split of four Amazon e-commerce datasets.}
\label{tab:train_stats}
\vspace{-20pt}
\end{table}

\section{Fairness Objectives}
\label{sec:objective}
In this section, we formally define the problem of fairness-aware explainable recommendation, considering both group and individual level fairness.
The group unfairness measurement over the entire population can be viewed as a between-group unfairness component.
We approximate such parity across a divided set of subgroups of the population, e.g., in accordance with 
user visibility.
The individual unfairness component is computed as an average sum of inequality in benefits received by individuals overall. We also formalize similar metrics in terms of individual-level fairness. 

\subsection{Group Unfairness Metrics}
Group fairness 
holds when users from two groups maintain the same probability of a positive decision \cite{dwork2018group}. 
In our setting, we consider a group of \emph{active} users $G_1$ vs.\ a  group of \emph{inactive} users $G_2$, defined according to the number of purchased items from historical records,
such that $G_1 \cap G_2 = \varnothing$.

Suppose there are $m$ users $\{u_1, u_2, \cdots, u_m\}$ associated with top-$N$ recommended items $\{v_1, v_2, \cdots, v_N\ \mid u_i \}$. In the following part, we take $\{1\leq i \leq m\}$ and $\{1\leq j \leq N\}$ to index users and items. We use $\mathbf{Q} = [Q_{ij}]_{m \times N}$, where $Q_{ij}\in\{0,1\}$ denotes whether item $j$ is selected for recommendation to user $u_i$. Let $\mathbf{Q_{i}} = [Q_{i1}Q_{i2} \cdots Q_{iN}]^{\mathsf{T}}$ represent the selection vector for the top-$K$ recommendation list of user $u_i$ under the constraints $\sum_{j=1}^{N}Q_{ij} = K$, $K \leq N$. We use the notation $\mathcal{F}$ to refer to a metric that scores the recommendation quality such that $\mathcal{F}(\mathbf{Q_{i}})$ denotes the recommendation quality for user $u_i$, invoking a metric such as $NDCG@K$ or F$_1$ score.

The group recommendation unfairness is defined as follows:
\begin{definition}{Group Recommendation Unfairness:}\label{def:GRU}
\begin{align}
GRU(G_1, G_2, \mathbf{Q}) = \left| \frac{1}{|G_1|}\sum_{i \in G_1} \mathcal{F}(\mathbf{Q_{i}}) - \frac{1}{|G_2|}\sum_{i \in G_2} \mathcal{F}(\mathbf{Q_{i}}) \right|
\end{align}
\end{definition}
As we have discussed in the previous section, it is the disparity in the diversity of the path distribution which leads to the performance disparity. In explainable recommendation over KGs, we also define group-level unfairness of the explanation path diversity by applying Simpson’s Index of Diversity ($SID$) to the user--item path distribution. In this context, we define:
\begin{definition}{Group Explanation Diversity Unfairness:}\label{def:GEDU}
\begin{align}
 GEDU(G_1, G_2, \mathbf{Q}) =\left| \frac{1}{|G_1|}\sum_{i \in G_1} f(\mathbf{Q_{i}}) - \frac{1}{|G_2|}\sum_{i \in G_2} f(\mathbf{Q_{i}}) \right|,
\end{align}\label{eq:GRU}
\end{definition}
\noindent where $f(\mathbf{Q_{i}})$ reflects the explanation fairness score in terms of the diversity of historical interactions of user $u_i$ with explainable paths (defined later in  Eq.~\ref{eq:path_fairess_score}). We adopt the averaging strategy to represent the user--item explanation diversity of two groups. 

\subsection{Individual Unfairness Metrics}
\label{sec:individual_metric}
The concept of individual fairness was first introduced by  \citet{dwork2012fairness} to address the inability of group fairness to reflect individual merits. The underlying notion is that similar individuals ought to be treated similarly. In our recommendation setting, it is not possible to meet the strict criteria 
of individual fairness, since we focus on solving the algorithm bias, not the inherent data imbalance. However, we can follow this idea to measure the individual unfairness with regard to both recommendation performance and explanation diversity. 

For this, we invoke the \emph{Gini coefficient} \cite{gini1921measurement}, which 
is commonly used in sociology and other fields to measure the inequality dispersion. 
It ranges from 0 to 1, where 1 represents maximal inequality in the sense that one single person has all the income or consumption, and all others have none, while 0 means perfect equality, where everyone has the same income value.
In our setting, the Gini coefficient is adopted to quantify the individual recommendation performance unfairness as follows:
\begin{definition}{Individual Recommendation Unfairness:}\label{def:IRU}
\begin{align}
 IRU(\mathbf{Q}) = \frac{\sum_{\mathbf{Q_{x}},\mathbf{Q_{y}}} \left| \mathcal{F}(\mathbf{Q_{x}}) -  \mathcal{F}(\mathbf{Q_{y}}) \right|} {2m\sum_{i=1}^{m} \mathcal{F}(\mathbf{Q_{i}})},
\end{align}
\end{definition}
\noindent where $x \neq y$ denotes two random users. Similarly, considering the explanation diversity in terms of fairness, we also define a measure of explanation diversity disparity among different individual users:

\begin{definition}{Individual Explanation Diversity Unfairness:} \label{def:IEDU}
\begin{align}
 IEDU(\mathbf{Q}) = \frac{\sum_{\mathbf{Q_{x}},\mathbf{Q_{y}}} \left| f(\mathbf{Q_{x}}) -  f(\mathbf{Q_{y}}) \right|} {2m\sum_{i=1}^{m} f(\mathbf{Q_{i}})}
\end{align}
\end{definition}

\subsection{Problem Formulation}
We can now proceed to formalize the problem of explainable recommendation under fairness constraints.
The original KG-based explainable recommendation problem aims to recover missing user--item interactions from an incomplete KG along with a set of paths that serve as the corresponding explanations for the recommendations.
Instead of imposing fairness constraints in the path-finding process, which would require substantial computational effort to retrain existing models, we seek a fairness-aware path reranking for explainable recommendation. Given a candidate set of user--item paths selected by an existing model, our goal is to rank the paths to obtain high-quality recommendations while satisfying fairness constraints.
Formally, we define the problem 
as follows.
\begin{definition}{\textbf{Group Fairness-aware Explainable Recommendation}}
Given a set of users from different groups $G_1, G_2$, each user $u$ having a set of user--item paths, and an integer $K$, the goal is to maximize the overall top $K$ recommendation quality by ranking these paths for each user under the fairness constraints that $GRU(G_1,G_2,\hat{\mathbf{Q}})\le\varepsilon_1$ and $GEDU(G_1,G_2,\hat{\mathbf{Q}})\le\varepsilon_2$,
where $\hat{\mathbf{Q}}\in\{0,1\}^{m\times N}$ and $\hat{\mathbf{Q}}_{ij}=1$ implying the $j$-th candidate path is selected among top $K$ outputs for the $i$-th user by the algorithm.
\end{definition}
Similarly, we can define the task of individual fairness-aware explainable recommendation by 
treating each user as belonging to a group and using  Eq.~\ref{def:IRU} and Eq.~\ref{def:IEDU} in the fairness constraints.
One important application is to quantify the unfairness of an algorithm, particularly assessing how different the outcomes are between different groups when using current state-of-the-art algorithms. Thus, we will attempt to minimize the unfairness while retaining the recommendation quality to the extent possible. 

\section{Fairness-Aware Algorithm}

\begin{table*}[!tp]
\centering
\small
\renewcommand\arraystretch{1.1}
\setlength{\tabcolsep}{3.5pt}
\begin{tabular}{c|cccccccc|cccccccc}
\cline{1-17}
\cline{1-17}
\cline{1-17}
\cline{1-17}
\multicolumn{1}{c|}{Dataset} & \multicolumn{8}{c|}{\textbf{CDs \& Vinyl}} & \multicolumn{8}{c}{\textbf{Clothing}} \\
\cline{1-17}
\multirow{2}{*}{Measures ($\%$)} & \multicolumn{2}{c}{Overall} & \multicolumn{2}{c}{Inactive Users} &  \multicolumn{2}{c}{Active Users} & \multicolumn{2}{c|}{GRU}  & \multicolumn{2}{c}{Overall}  & \multicolumn{2}{c}{Inactive Users} & \multicolumn{2}{c}{Active Users} & \multicolumn{2}{c}{GRU}  \\
 & NDCG & F$_1$ & NDCG & F$_1$ &  NDCG  &  F$_1$ &  NDCG  &  F$_1$ & NDCG & F$_1$ & NDCG & F$_1$ &  NDCG  &  F$_1$ &  NDCG &  F$_1$ \\
\cline{1-17}
HeteroEmbed   & 6.992 & 3.576 & 6.526 & 3.373 & 15.843 & 7.429 & 9.317 & 4.056 & 3.221  & 1.404 & 3.121 & 1.348 & \textbf{5.130} & 2.461  & 2.009 & 1.113 \\
Fair HeteroEmbed       & \textbf{8.094} & \textbf{4.019} & \textbf{7.674} & \textbf{3.820} & \textbf{16.074} & \textbf{7.801} & \textbf{8.400} & \textbf{3.981} & \textbf{3.494} & \textbf{1.536} & \textbf{3.484} & \textbf{1.482} & 3.691 & \textbf{2.556} & \textbf{0.207} & \textbf{1.074}  \\
\cline{1-17}
PGPR        & 6.947 & 3.571 & 6.526 & 3.373 & 14.943 & 7.324 & 8.417 & 3.951 & 2.856 &1.240 & 2.787 & 1.198 & \textbf{4.197} & 2.036  & 1.410 & 0.833 \\
Fair PGPR      & \textbf{8.045} &  \textbf{4.019} &  \textbf{7.675} & \textbf{3.820} & \textbf{15.074} & \textbf{7.801} & \textbf{7.399} & \textbf{3.261}  & \textbf{3.101} & \textbf{1.314}  & \textbf{3.089} & \textbf{1.274} & 3.322 & \textbf{2.078} & \textbf{0.233} & \textbf{0.804}  \\
\cline{1-17}
KGAT        & 5.411 & 3.357 & 5.038 & 3.162 & 12.498 & 7.046 & 7.460 & 3.884 & 3.021 & 1.305 & 2.931 & 1.254 &  4.741 & 2.259  & 1.810 & 1.005 \\
Fair KGAT    & \textbf{5.640} & \textbf{3.492} & \textbf{5.295} & \textbf{3.318} & 12.366 & 6.791 & \textbf{7.081} & \textbf{3.473} & \textbf{3.206} & \textbf{1.393} & \textbf{3.119} & \textbf{1.347} &  \textbf{4.843} & \textbf{2.262}  & \textbf{1.724} & \textbf{0.915}  \\
\cline{1-17}
\cline{1-17}
\cline{1-17}
\cline{1-17}
Dataset & \multicolumn{8}{c|}{\textbf{Beauty}} & \multicolumn{8}{c}{\textbf{Cell Phones}} \\
\cline{1-17}
\multirow{2}{*}{Measures ($\%$)}
& \multicolumn{2}{c}{Overall} & \multicolumn{2}{c}{Inactive Users} &  \multicolumn{2}{c}{Active Users} & \multicolumn{2}{c|}{GRU}  & \multicolumn{2}{c}{Overall}  & \multicolumn{2}{c}{Inactive Users} & \multicolumn{2}{c}{Active Users} & \multicolumn{2}{c}{GRU}  \\
 & NDCG & F$_1$ & NDCG & F$_1$ &  NDCG  &  F$_1$ &  NDCG  &  F$_1$ & NDCG & F$_1$ & NDCG & F$_1$ &  NDCG  &  F$_1$ &  NDCG &  F$_1$ \\
\cline{1-17}
HeteroEmbed       & 6.371 & 3.125 & 6.078 & 2.756 & 11.933 & \textbf{10.132} & 5.855 & 7.376 & 5.833 & 2.537 & 5.645 & 2.311 & \textbf{9.395} & \textbf{6.829}  & 3.750 & 4.518\\
Fair HeteroEmbed  & \textbf{6.740} & \textbf{3.181} & \textbf{6.451} & \textbf{2.924} & \textbf{12.229} & 9.853 & \textbf{5.778} & \textbf{6.929} & \textbf{6.199} & \textbf{2.678} & \textbf{6.037} & \textbf{2.466} & 9.284 & 6.648 & \textbf{3.247} & \textbf{4.182}  \\
\cline{1-17}
PGPR              & 5.456 & 2.544 & 5.219 & 2.291 & 9.766 & 7.349 & 4.547 & 5.148 & 5.079 & 2.116 & 4.945 & 1.972 & 7.626 & \textbf{4.846} & 2.681 &  2.874 \\
Fair PGPR         & \textbf{5.717} & \textbf{2.680} & \textbf{5.504} & \textbf{2.430} & \textbf{9.766} & \textbf{7.431} & \textbf{4.262} & \textbf{5.001} & \textbf{5.380} & \textbf{2.193} & \textbf{5.252} & \textbf{2.076} & \textbf{7.807} & 4.409 & \textbf{2.555} & \textbf{2.333} \\
\cline{1-17}
KGAT       & 6.108 & 3.169 & 5.863 & 2.761 & 10.763 & 10.929 & 4.900 & 8.168 & 5.111 & 2.265 & 4.958 & 2.100 &  8.026 & 5.391 & 3.068 & 3.297\\
Fair KGAT  & \textbf{6.241} & \textbf{3.228} & \textbf{6.001} & \textbf{2.832} & \textbf{10.785} & 10.752 & \textbf{4.784} & \textbf{7.920} & \textbf{5.304} & \textbf{2.391} & \textbf{5.159} & \textbf{2.240} &  \textbf{8.057} & 5.256 & \textbf{2.898} & \textbf{3.016}  \\
\cline{1-17}
\cline{1-17}
\cline{1-17}
\cline{1-17}
\end{tabular}
\caption{Overall recommendation performance of inactive users and active users of our proposed fairness-aware algorithm on explainable recommendation approaches and corresponding baselines on four Amazon datasets. The results are reported in percentage (\%) and are calculated based on the top-10 predictions in the test set. The best results are highlighted in bold. HeteroEmbed is proposed in \cite{ai2018learning} and PGPR, KGAT come from \cite{xian2019kgrl} and \cite{Wang2019KGATKG}, repectively.} 
\label{tab:eval}
\end{table*}
\label{sec:alg}
In this paper, we propose a general fairness-aware ranking framework, which can be applied on top of several state-of-the-art explainable recommendation algorithms based on knowledge graphs. With an understanding of the fairness goals, the question that arises is: How could an explainable recommendation model be able to yield a fair  recommendation list while maintaining explanation diversity? 
The original explainable paths provided by the original algorithms
provide compelling arguments for why a given item is recommended based on historical user interactions. It would not make sense to completely ignore all of them. At the same time, fairness constraints raise path diversity as an additional concern.
On the basis of this, our fairness-aware algorithm should consider both historical user interactions and the diversity of generated explainable paths.

We introduce two fairness scores: the \emph{path score} $\mathcal{S}_\textrm{p}$ and the \emph{diversity score} $\mathcal{S}_\textrm{d}$.
\paragraph{Path Score.}
The path score weights the quality of paths.
According to the motivation of fairness concerns in Sec.~\ref{sec:fair_concern}, we wish to consider a more varied set of paths rather than just the kinds that dominate the historic user--item interaction data. 
Therefore, our path score incorporates an explicit debiasing weighting to adjust the bias of user--item path patterns in historical records:
\begin{align}\label{eq:penalization_score}
\mathcal{S}_{\textrm{p}}(\mathbf{Q_{i}}) &= \sum_{j=1}^{N} \sum_{\pi \in \Pi} Q_{ij} \mathcal{S}{\pi}(i,\pi),
\end{align}
\noindent where $\mathbf{Q_{i}}$ is the vector of recommended items as defined in Sec.~\ref{sec:objective}. 
For a path pattern $\pi$ and a user $u$, we use $\mathcal{S}(u,\pi)$ as the coefficient expressing the preference adjustment for the user $u_i$.
\begin{align}\label{eq:penalization_score_detail}
\mathcal{S}(u,\pi) &= \frac{w_{\pi} }{\mathcal{D}_{u,\mathcal{V}}(\pi)}  \sum_{L_{uv}^\pi\in\mathcal{L}_{uv}^\pi}  \ell_\mathrm{path} \left(L_{uv}^\pi \right)
\end{align}
Here,  
$\mathcal{L}_{uv}^\pi$ is the set of all positive user--item paths starting from user $u$ with respect to path pattern $\pi$ to item $v$. 
$\ell_\mathrm{path}(L_{uv}^\pi) = \sum_{i=1}^{|\pi|}(\vec e_0 + \vec r_i) \cdot \vec e_i$ (where $e_0=u$ and $e_{|\pi|}=v$) is the score of a path $L_{uv}^\pi=\{u, r_1, e_1, r_2, e_2, \ldots, \allowbreak e_{|\pi|-1},r_{|\pi|}, v\}$, which can easily be computed following the methods proposed for current KG-based explainable recommendation systems.
$w_{\pi}$ represents the weight of the path pattern $\pi$ in generated explainable paths:
\begin{align}
w_{\pi} &= \log\left(2 + \frac{ |\mathcal{L}_{uv}^\pi |}{ \sum_{\pi'} \mid \mathcal{L}_{uv}^{\pi'}|} \right)
\end{align}
Finally,
$\mathcal{D}_{u,\mathcal{V}}(\pi)$ is the weight of path pattern $\pi$ in the training set defined in Sec.~\ref{sec:pre}. On the one hand, we retain the existing weights of path scores as $w_{\pi}$, while on the other hand striving to break the intrinsic explainable path patterns assigned to different groups of users. For the latter, $\mathcal{D}_{u,\mathcal{V}}(\pi)$ serves as a regularization factor to minimize the path diversity bias between the groups.

\paragraph{Diversity Score.}
Additionally, we also consider the fairness with regard to explainable path diversity. For each user--item pair $(u, v)$, we get access to the user--item path distribution based on the retrieved explainable paths of the original algorithm \cite{ai2018learning, xian2019kgrl}, as 
$\{|\mathcal{L}_{uv}^{\pi_1}|, |\mathcal{L}_{uv}^{\pi_2}|, \allowbreak\ldots, |\mathcal{L}_{uv}^{\pi_{|\Pi|}}\}$. Let $\Pi$ be a number of valid user--item path patterns in the KG. 
Then we are able to calculate Simpson's Index of Diversity of such $(u, v)$ pairs as $\mathcal{T}_{uv}(\Pi,N_v)$, 
as defined in Sec.~\ref{sec:diversity_simpson}, where $N_v$ denotes the total number of retrieved paths starting from user $u$ and ending up at item $v$ as $\sum_{\pi'} \mid \mathcal{L}_{uv}^{\pi'}|$. We can then define our diversity score, which can be regarded as introducing regularization for explainable path diversity:
\begin{align}\label{eq:diversity_score}
\mathcal{S}_\textrm{d}(\mathbf{Q_{i}}) &= \sum_{j=1}^{N} \sum_{|\pi|} Q_{ij} \mathcal{T}_{ij}(\Pi,N_j)
\end{align}
\paragraph{Fairness score.}
By aggregating the personalization and diversity scores, we can calculate the fairness score for user $u_i$ defined as:
\begin{align}\label{eq:path_fairess_score}
    f_i(\mathbf{Q_{i}}) = \alpha \mathcal{S}_\textrm{p}(\mathbf{Q_{i}}) + (1- \alpha)\lambda \mathcal{S}_\textrm{d}(\mathbf{Q_{i}}),
\end{align}
\noindent where $\alpha\in[0,1]$ is the weighting factor of path score $\mathcal{S}_\textrm{p}(\mathbf{Q_{i}})$, compared to the diversity score $\mathcal{S}_\textrm{d}(\mathbf{Q_{i}})$. $\lambda$ is a scaling factor so that $\mathcal{S}_\textrm{p}$ and $\mathcal{S}_\textrm{d}$ can be normalized onto the same scale.
\paragraph{Recommendation score}
We follow the baseline method of calculating the preference score $\mathcal{S}(i,j)$ between user $u_i$ and item $v_j$. Our goal becomes to find a selection strategy $\mathcal{R}_\textrm{rec}(\mathbf{Q_{i}}) = \sum_{j=1}^{N} Q_{ij} \mathcal{S}(i,j)$ for each user $u_i$ to recommend $K$ items that meet the group fairness constraints. Therefore, we can formulate the optimization recommendation problem as follows:
\begin{align}\label{eq:optimization}
\max_{Q_{ij}} \quad & \mathcal{R} = \sum_{i=1}^{m} \mathcal{R}_\textrm{rec}(\mathbf{Q_{i}})\\
\text{s.t.}\quad & \sum_{j=1}^{N} Q_{ij} = K, ~~Q_{ij} \in \{0,1\} \\
          & GRU(G_1, G_2, \mathbf{Q}) < \varepsilon_1 \label{eq:group_per}\\
          & GREU(G_1, G_2, \mathbf{Q}) < \varepsilon_2 \label{eq:group_div}
\end{align}
Here, Eqs.~\ref{eq:group_per} and \ref{eq:group_div} refer to the \emph{GRU} and \emph{GEDU} proposed in Def.~\ref{def:GRU} and  \ref{def:GEDU}, respectively,
with $\varepsilon_1$ and $\varepsilon_2$ representing recommendation performance and fairness disparity of baselines, correspondingly. Moreover, the optimization could also be extended to individual fairness constraints. We are able to take the constraints of the Gini coefficient difference as proposed in Defs.~\ref{def:IRU} and \ref{def:IEDU}. In this case, we replace Eqs.~\ref{eq:group_per} and \ref{eq:group_div} with following:
\begin{align}
& IRU(\mathbf{Q}) < \varepsilon_3 \label{eq:indi_per}\\
& IREU(\mathbf{Q}) < \varepsilon_4 \label{eq:indi_div}
\end{align}
Here, $\varepsilon_3$ and $\varepsilon_4$ are the individual recommendation and fairness disparity of the corresponding baselines.

The optimization of Eq.~\ref{eq:optimization} can be cast as a \emph{0-1 integer programming} optimization problem.
Although it is NP-complete, 
we can use fast heuristics\footnote{We use the 
Gurobi solver, \url{https://www.gurobi.com/}.} to find feasible solutions. 
While these may converge to a local optimum rather than a global one, our empirical findings show that the fairness-aware top-$K$ selection obtained is superior enough compared to the baseline methods. 
Further details will be given in Sec.~\ref{sec:exp}.

\paragraph{Ranking Score}
After solving the optimization problem and selecting \emph{which} items to recommend under fairness constraints for each user as $\mathbf{\hat Q_{i}}$, we still need to \emph{rank} the $K$ items to determine in which order they are presented. This also allows us to better compare the results against the baseline methods. 
Specifically, we create a top-k recommendation list with items ranked in descending order by the optimized recommendation score and the fairness score, defined as: 
\begin{align}
    \mathcal{R}_\textrm{rank}(\mathbf{\hat Q_{i}}) = \beta \gamma f_i(\mathbf{\hat Q_{i}}) + (1- \beta)
    \mathcal{R}_\textrm{rec}(\mathbf{\hat Q_{i}}),
    \label{eq:rank_fairess_socre}
\end{align}
where $\beta\in[0,1]$ is the weighting factor of the ranking fairness score. Similar to the $\lambda$ above, we also add a factor $\gamma$, which achieves the scaling effect.

\section{Experiments}
\label{sec:exp}
In this section, we first briefly describe the four real-world e-commerce datasets used for experiments.
Then, we evaluate our proposed fairness-aware algorithm on top of existing explainable recommendation approaches. A series of quantitative and qualitative analyses demonstrate the positive effects on both fairness metrics and the recommendation performance.

\subsection{Dataset and Experimental Setup}
\paragraph{Datasets}
All of our experiments are based on Amazon item e-commerce datasets \cite{he2016ups}. The collection consists of four different domains: \emph{CDs and Vinyl}, \emph{Clothing}, \emph{Cell Phones}, and \emph{Beauty}.
It should be noted that each dataset is considered as an independent benchmark that constitutes a respective knowledge graph with entities and relations crawled from Amazon\footnote{https://www.amazon.com/}. Thus, the evaluation results are not comparable over different domains.
The statistical details and train/test split 
correspond exactly to those of previous work in this area \cite{ai2018learning,xian2019kgrl}. Note that following  \citet{xian2019kgrl}, there is no constraint on possible path patterns, so any path between a user and the recommended item is considered to be valid. However, since shorter paths are more reliable for users as  explanation for recommendation, we only consider user--item paths with a length of up to $3$. We utilize the path patterns extracted from the KG dataset by the baselines for an equal comparison.

\paragraph{Experimental Setup}
As a prior investigation to verify the rationality of the unfairness, we first study the user interaction difference in terms of path distribution and the recommendation performance disparity over three state-of-art explainable RSs over KGs \cite{ai2018learning, xian2019kgrl,Wang2019KGATKG}. They rank the recommendation list by calculating the relevance scores of user--item pairs. \citet{ai2018learning} picks items using translational entity embeddings, while \citet{xian2019kgrl} computes the score as a path reward via reinforcement learning. \citet{Wang2019KGATKG} propose an attention-based collaborative knowledge graph method.

Following previous work, we consider as metrics the Normalized Discounted Cumulative Gain (\textbf{NDCG}) and \textbf{F$_1$} scores, two popular metrics to evaluate the recommendation performance. F$_1$ scores provide the harmonic mean of precision and recall, while NDCG evaluates the ranking by considering the position of correctly recommended items. 
We evaluate the group fairness and individual fairness in terms of the recommendation quality and explanation diversity, with the metric defined in Sec.~\ref{sec:objective}.

\begin{figure}[t]
\centering
\includegraphics[width=1.09in,height=0.85in]{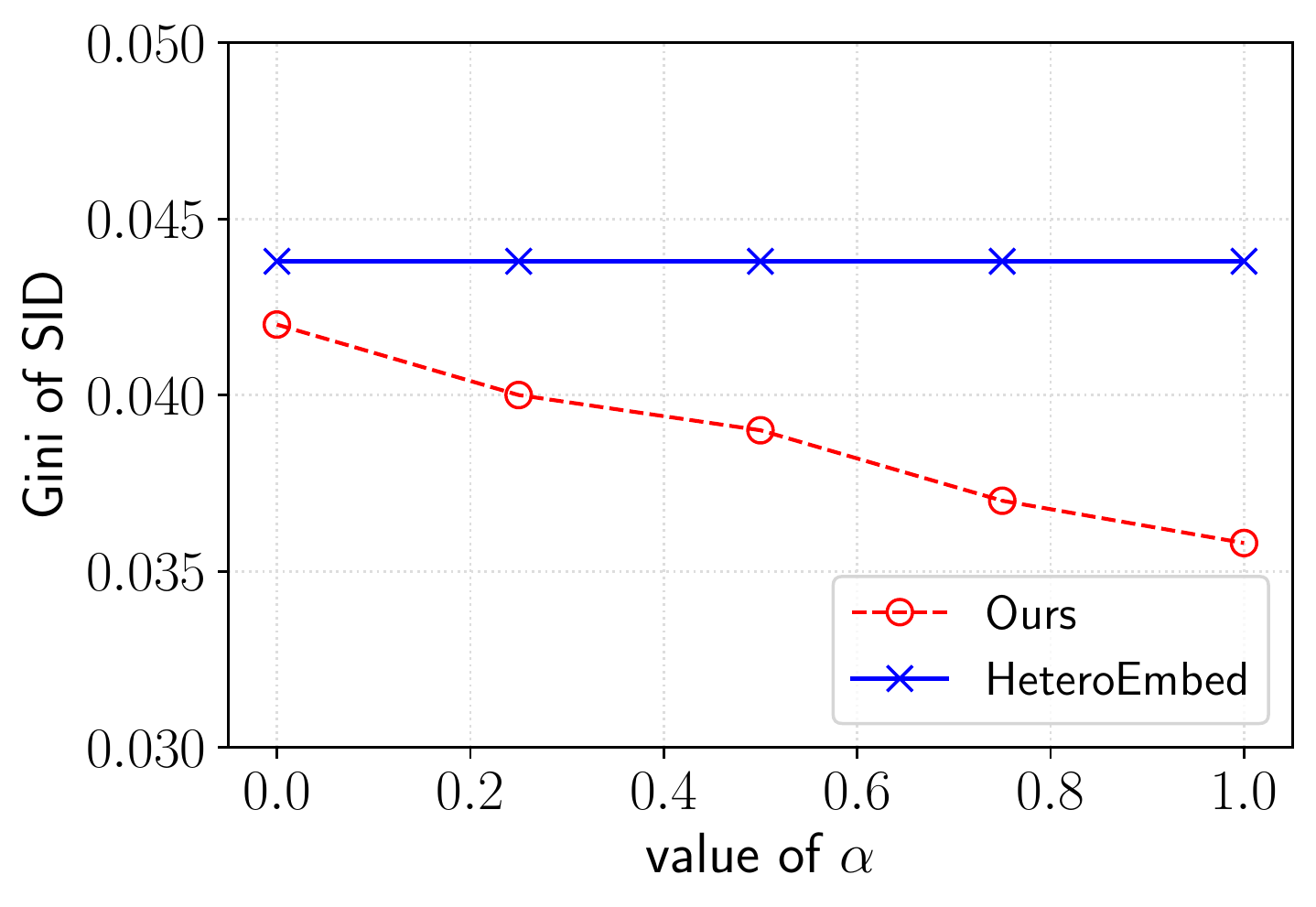}
\includegraphics[width=1.09in,height=0.85in]{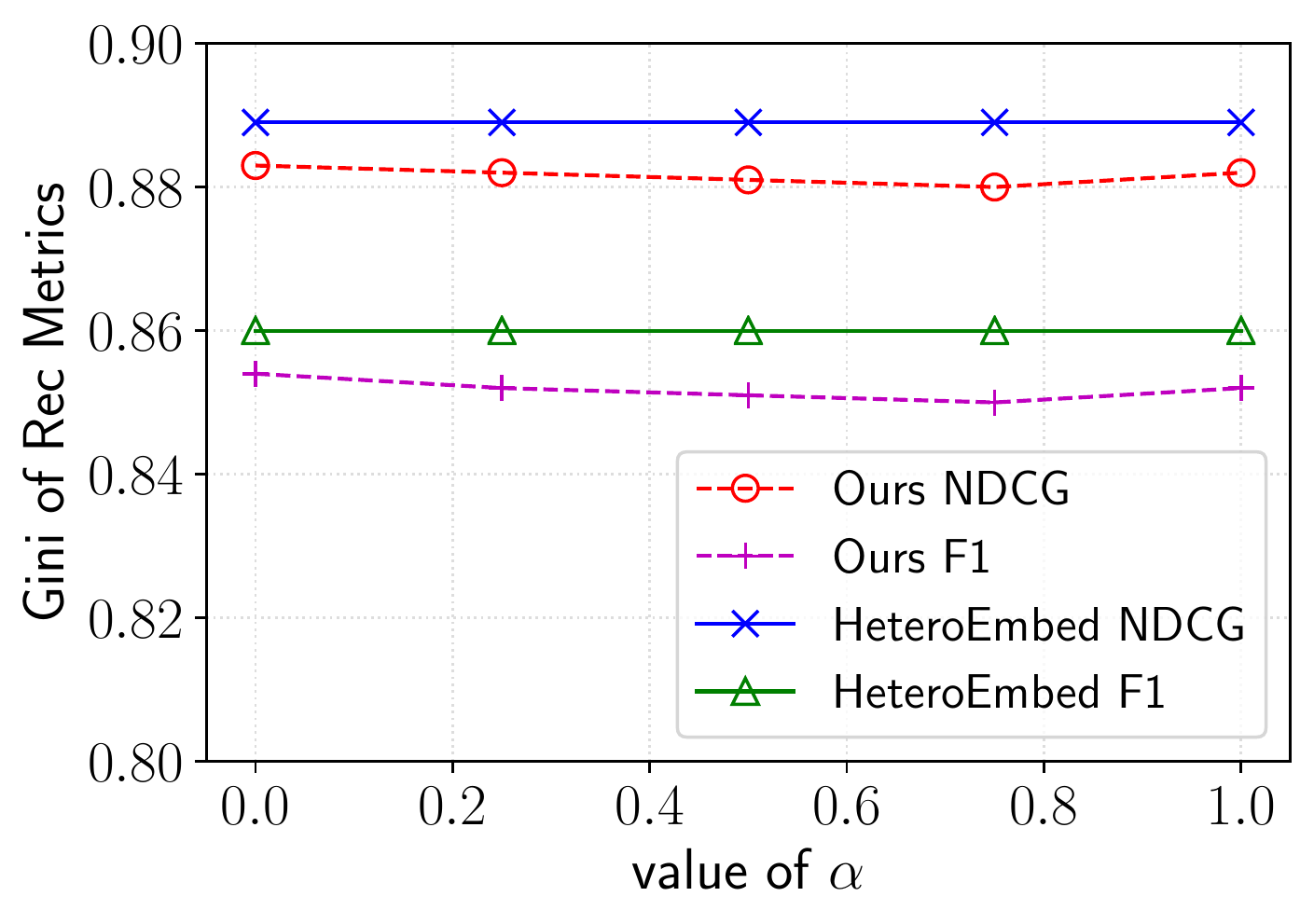}
\includegraphics[width=1.09in,height=0.85in]{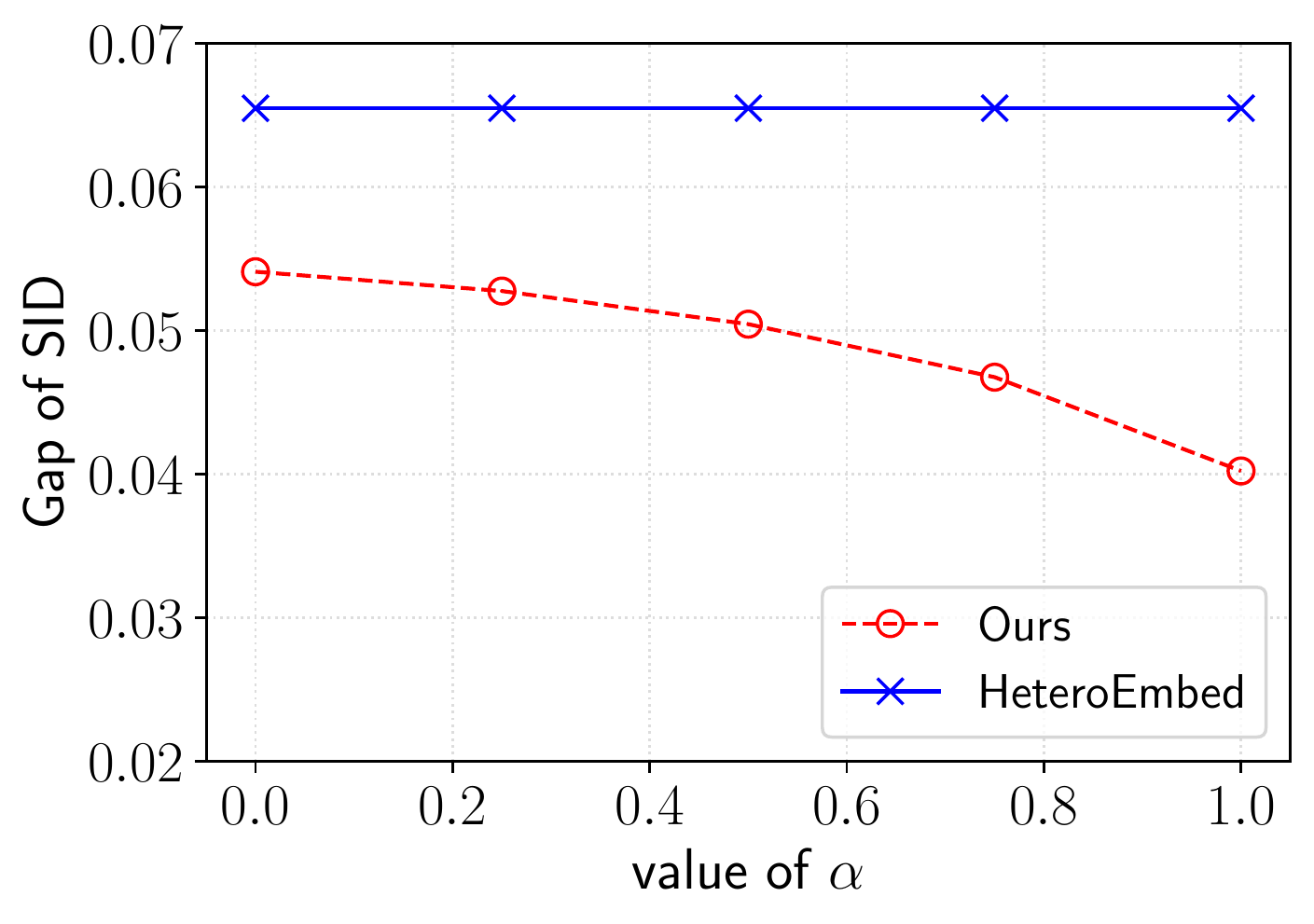} \\
\hspace*{0.2in} (a) Gini of $SID$ \hspace{0.15in} (b) Gini of Rec. Metrics  \hspace{0.1in} (c) Gap of $SID$
\vspace{3pt}
\includegraphics[width=1.09in,height=0.85in]{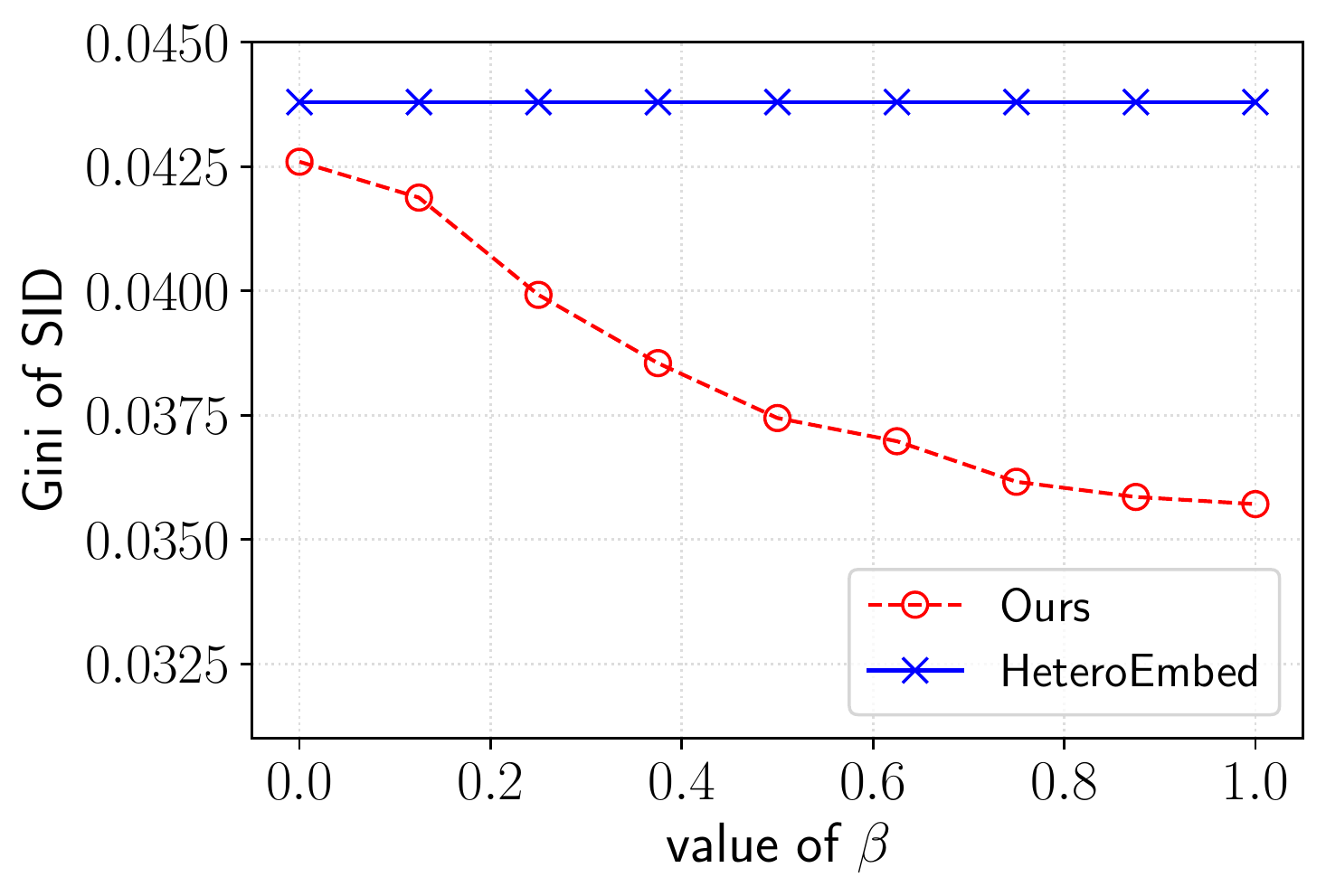}
\includegraphics[width=1.09in,height=0.85in]{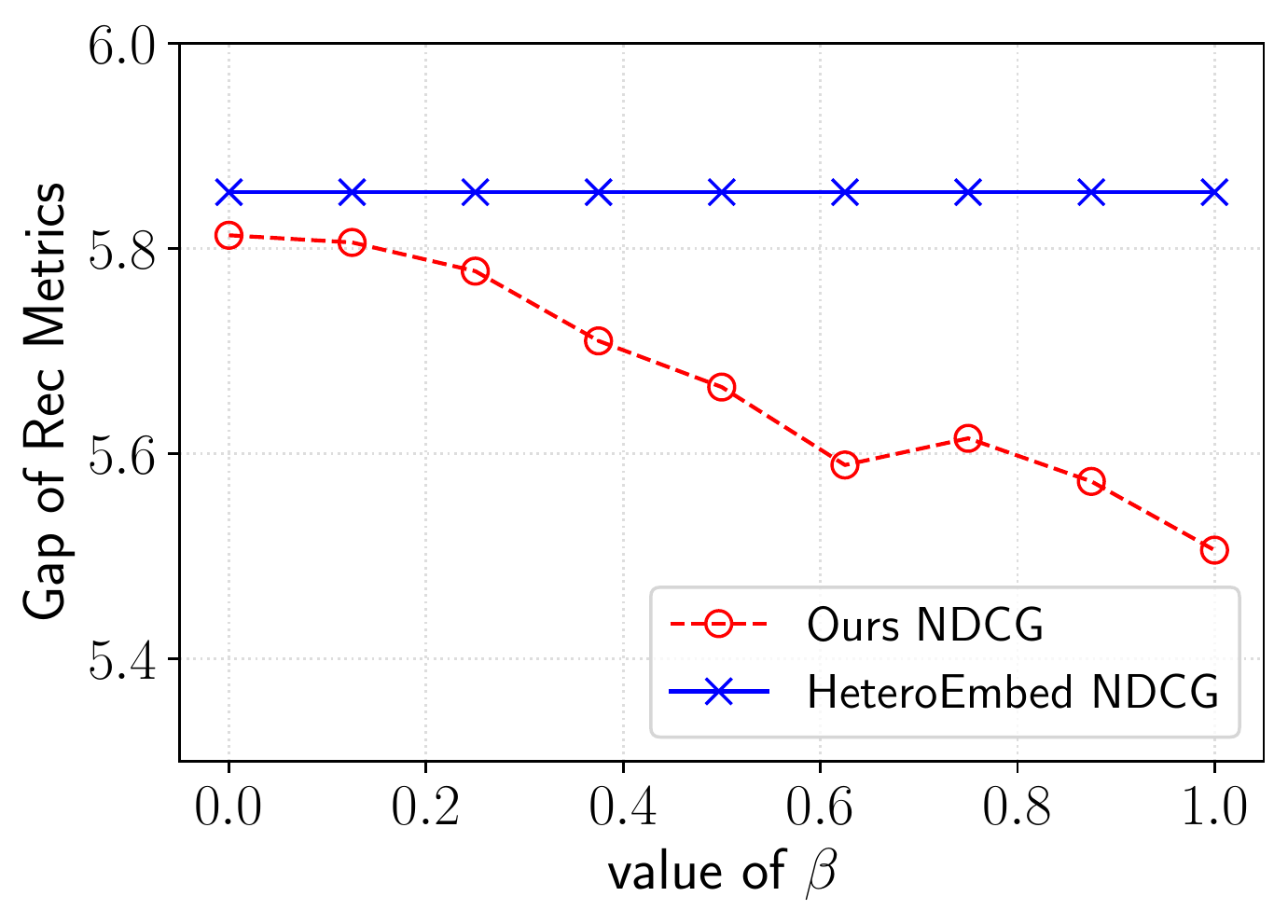}
\includegraphics[width=1.09in,height=0.85in]{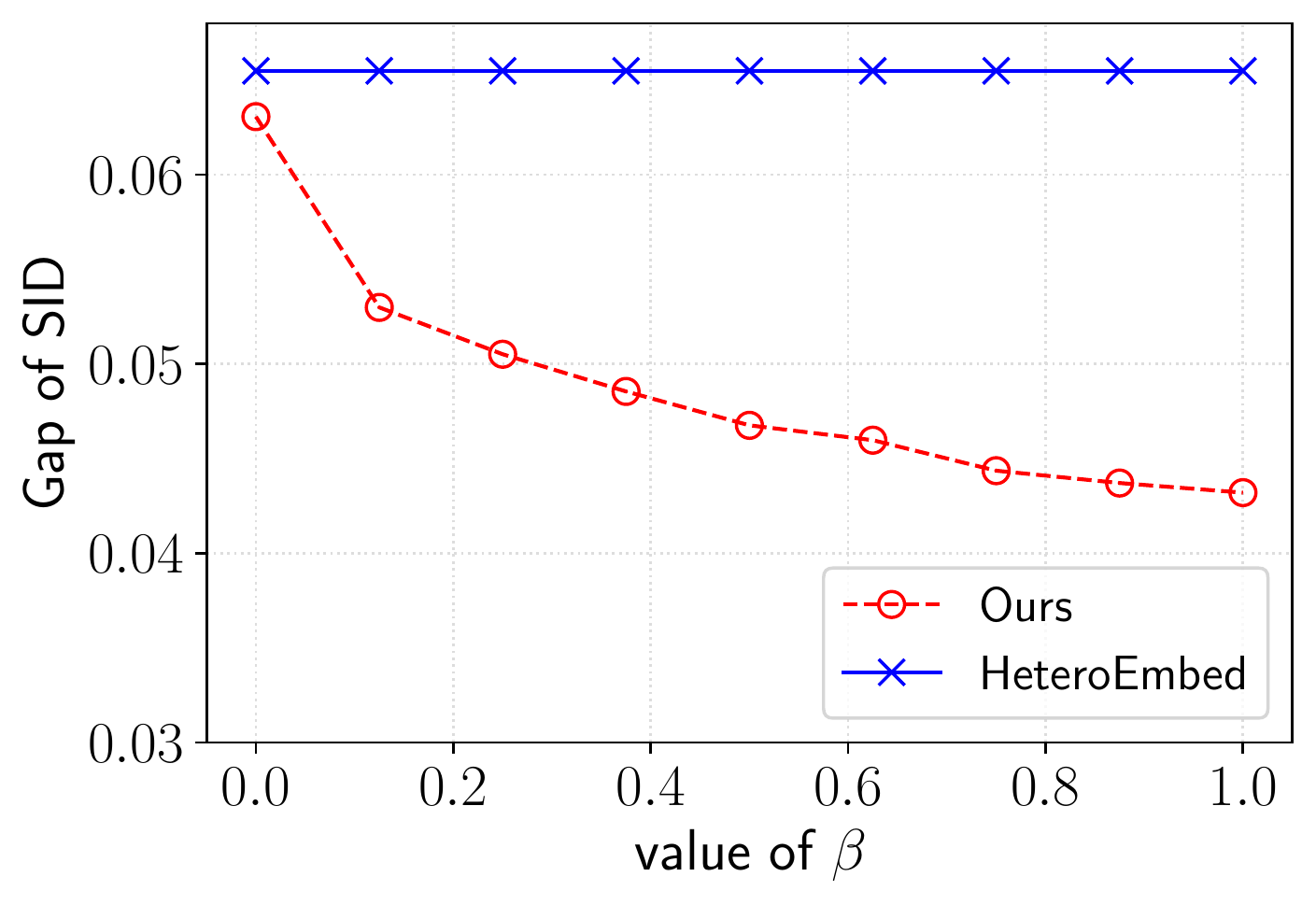} \\
\hspace*{-0.05in} (d) Gini of $SID$ \hspace{0.15in} (e) Gap of Rec. Metrics  \hspace{0.4in} (f) Gap of $SID$
\vspace{-5pt}
\caption{Results of Fairness-aware HeteroEmbed on Beauty dataset when fix $\beta=0.5$ (a-c) and fix $\alpha=0.75$ (d-f).}
\label{fig:HeteroEmbed_beauty}
\vspace{-5pt}
\end{figure}
\begin{figure}[t]
\centering
\includegraphics[width=1.09in,height=0.85in]{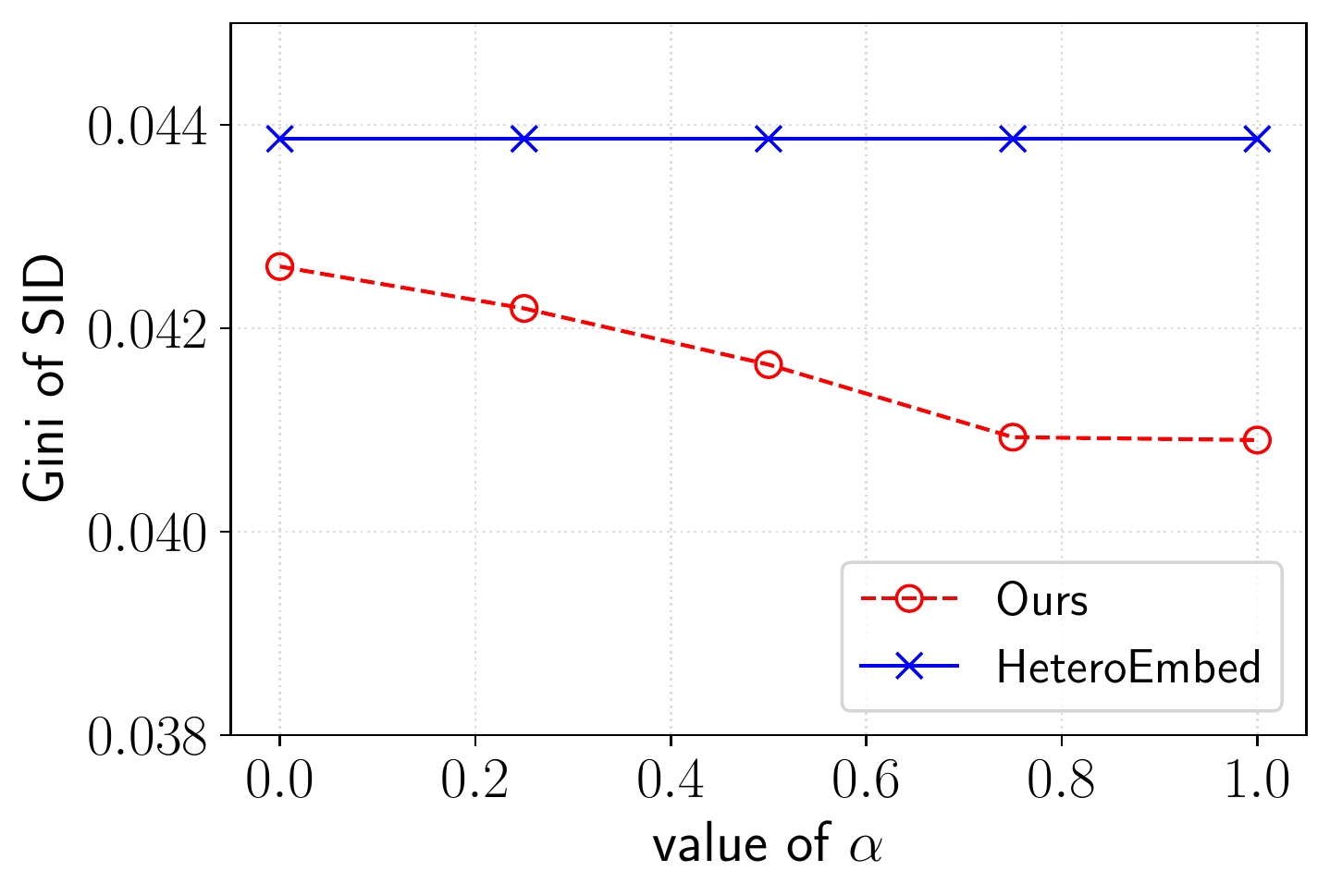}
\includegraphics[width=1.09in,height=0.85in]{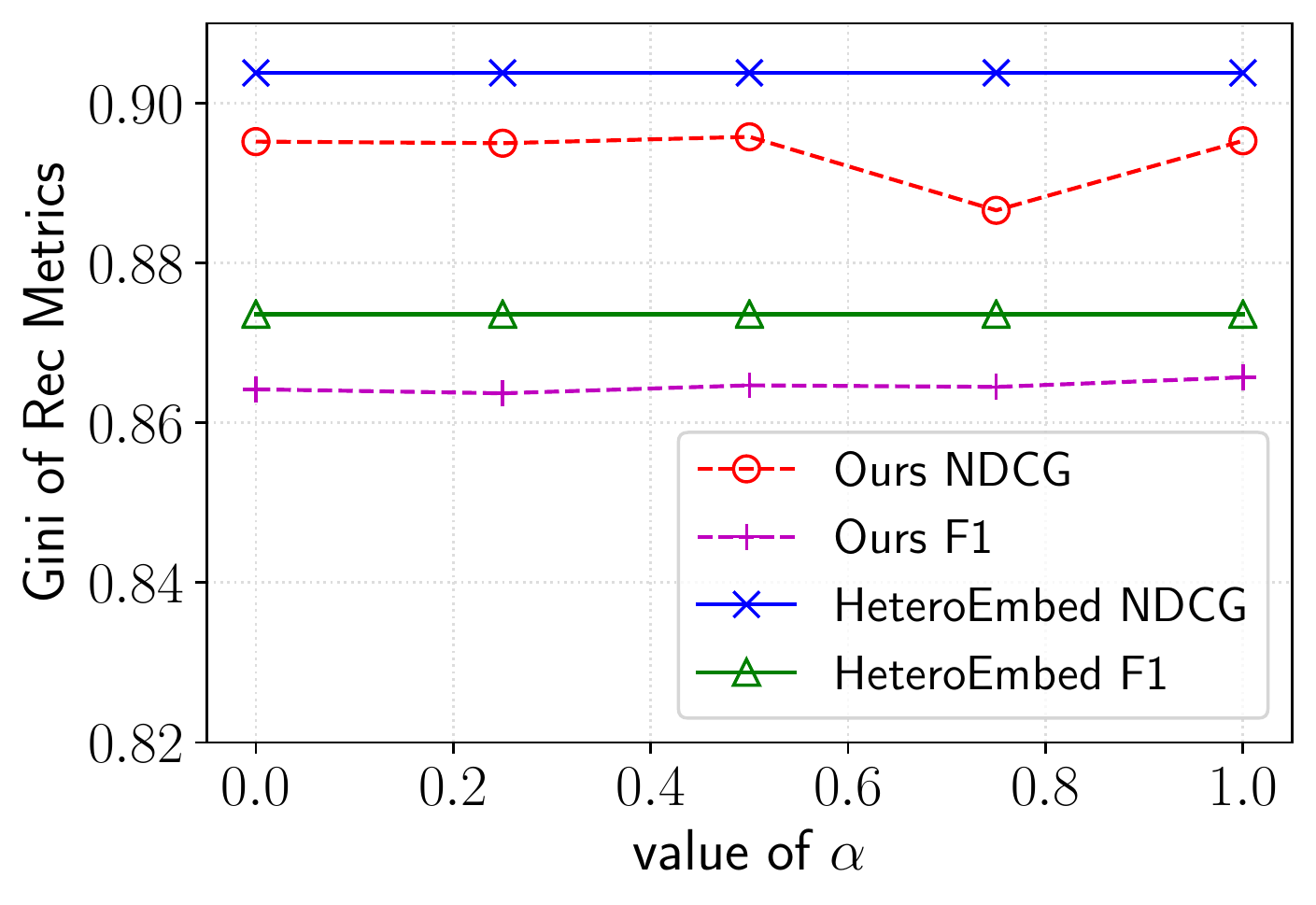}
\includegraphics[width=1.09in,height=0.85in]{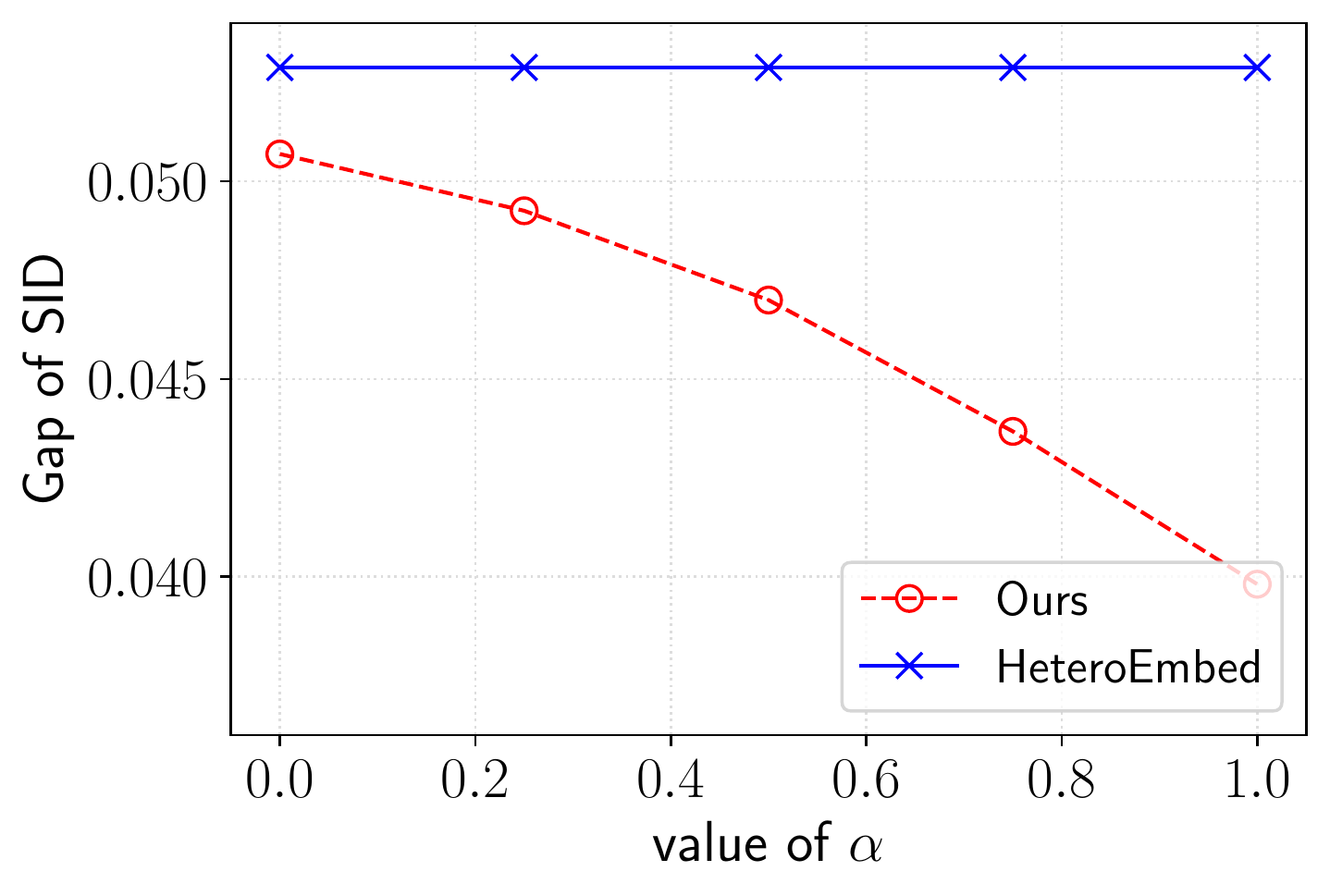} \\
\hspace*{0.2in} (a) Gini of $SID$ \hspace{0.15in} (b) Gini of Rec. Metrics  \hspace{0.1in} (c) Gap of $SID$
\vspace{3pt}
\includegraphics[width=1.09in,height=0.85in]{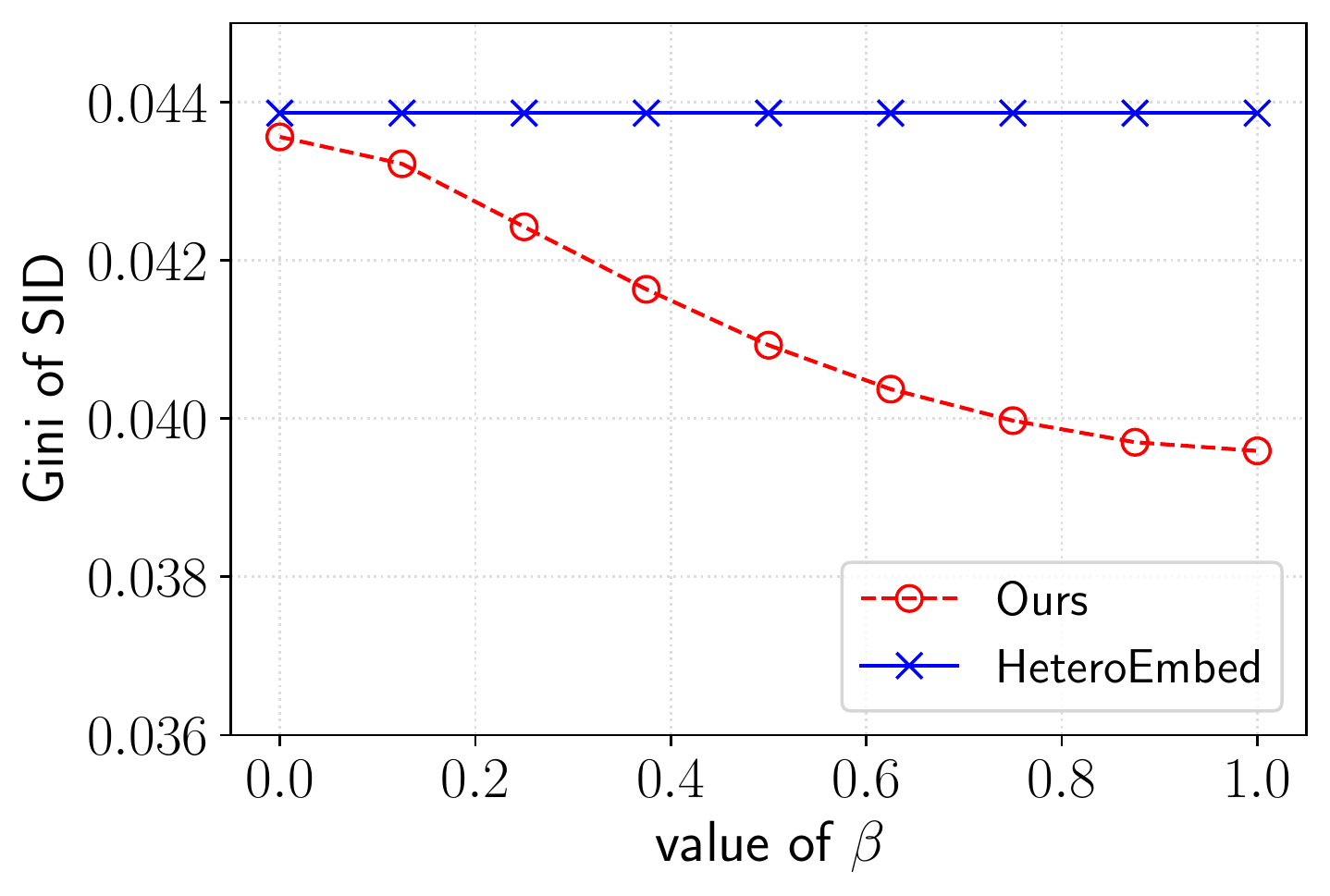}
\includegraphics[width=1.09in,height=0.85in]{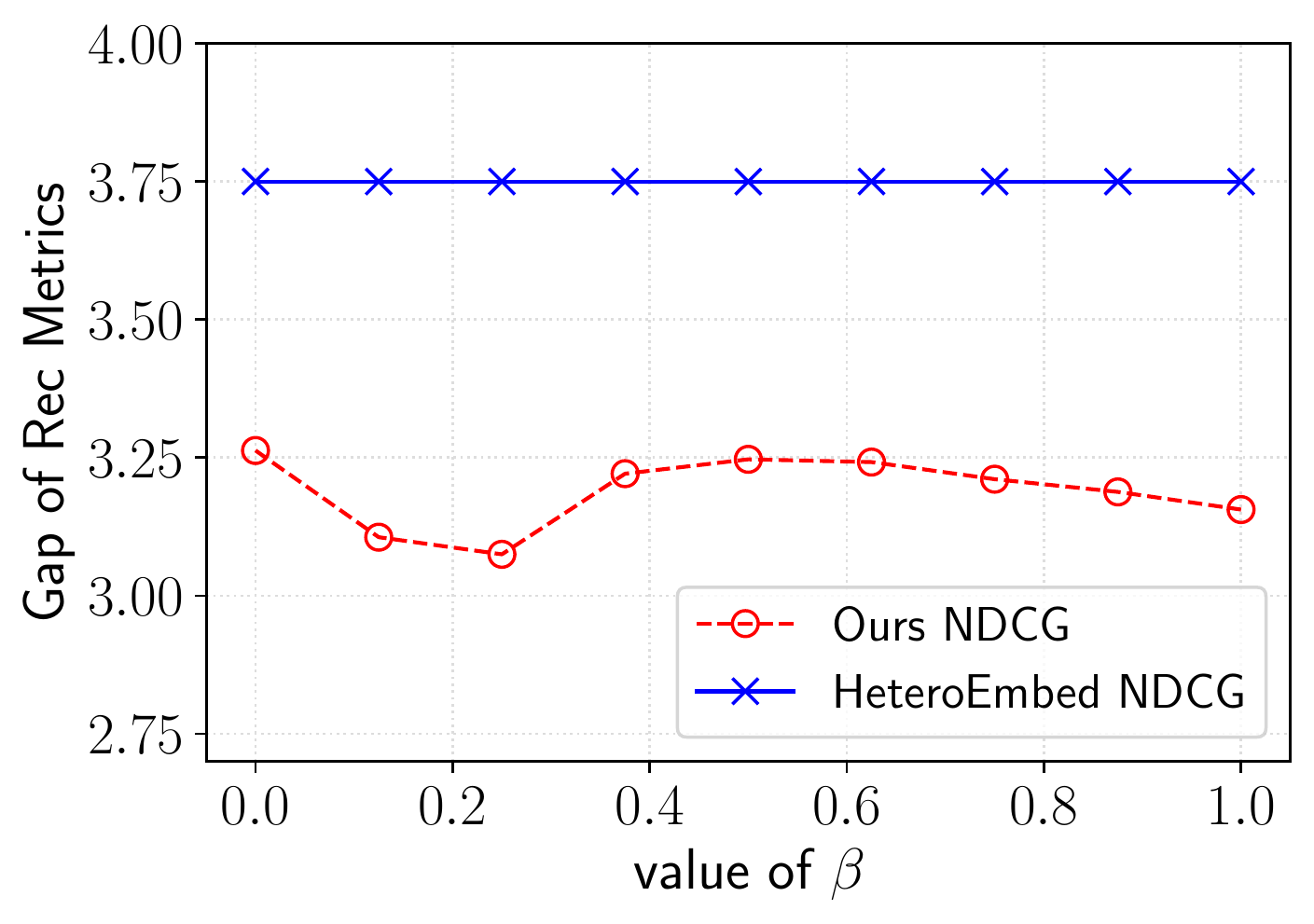}
\includegraphics[width=1.09in,height=0.85in]{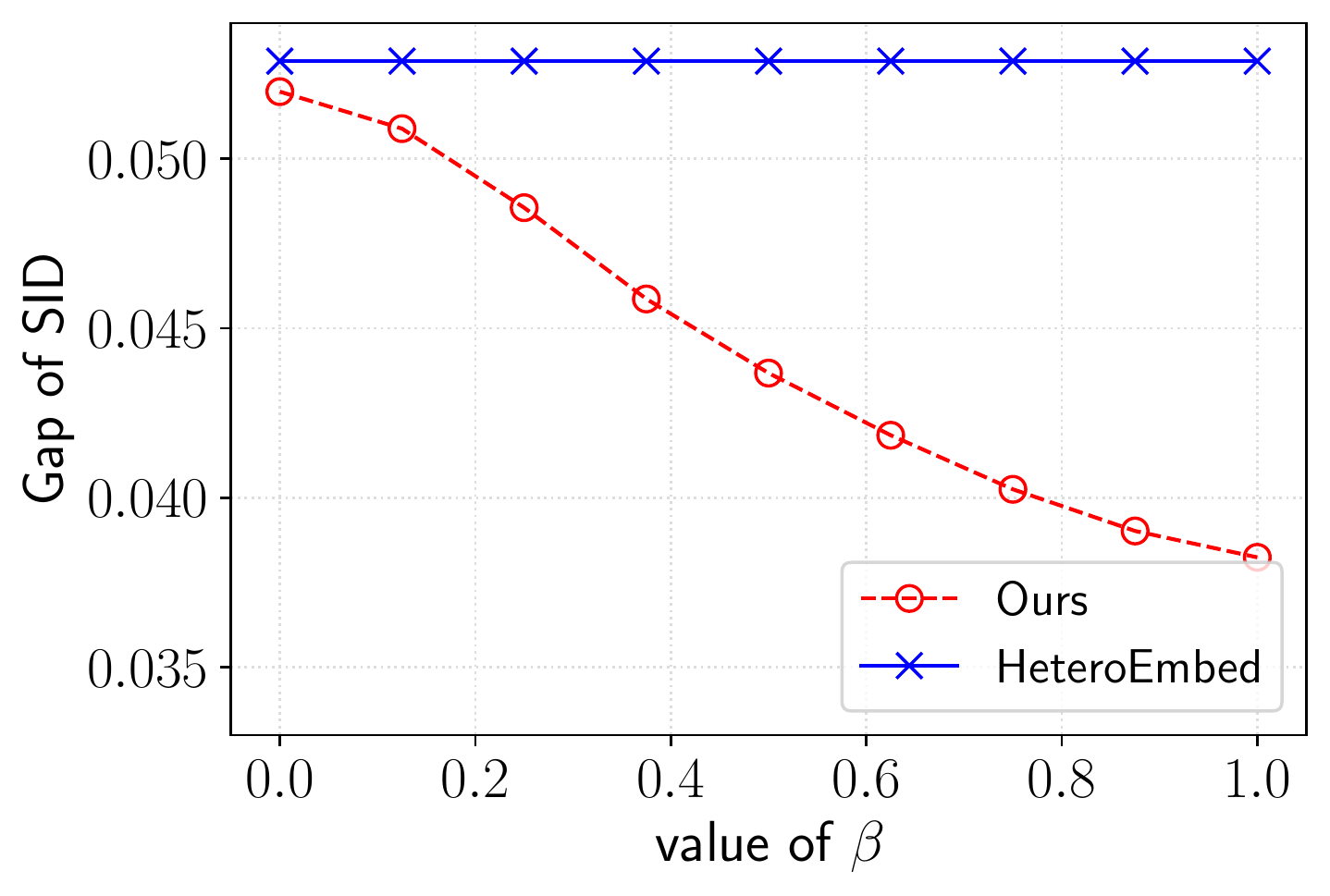} \\
\hspace*{-0.05in} (d) Gini of $SID$ \hspace{0.15in} (e) Gap of Rec. Metrics  \hspace{0.4in} (f) Gap of $SID$
\vspace{-5pt}
\caption{Results of Fairness-aware HeteroEmbed on Cell Phone dataset when fix $\beta=0.5$ (a-c) and fix $\alpha=0.75$ (d-f).}
\label{fig:HeteroEmbed_cell}
\vspace{-5pt}
\end{figure}
\begin{figure}[t]
\centering
\includegraphics[width=1.09in,height=0.85in]{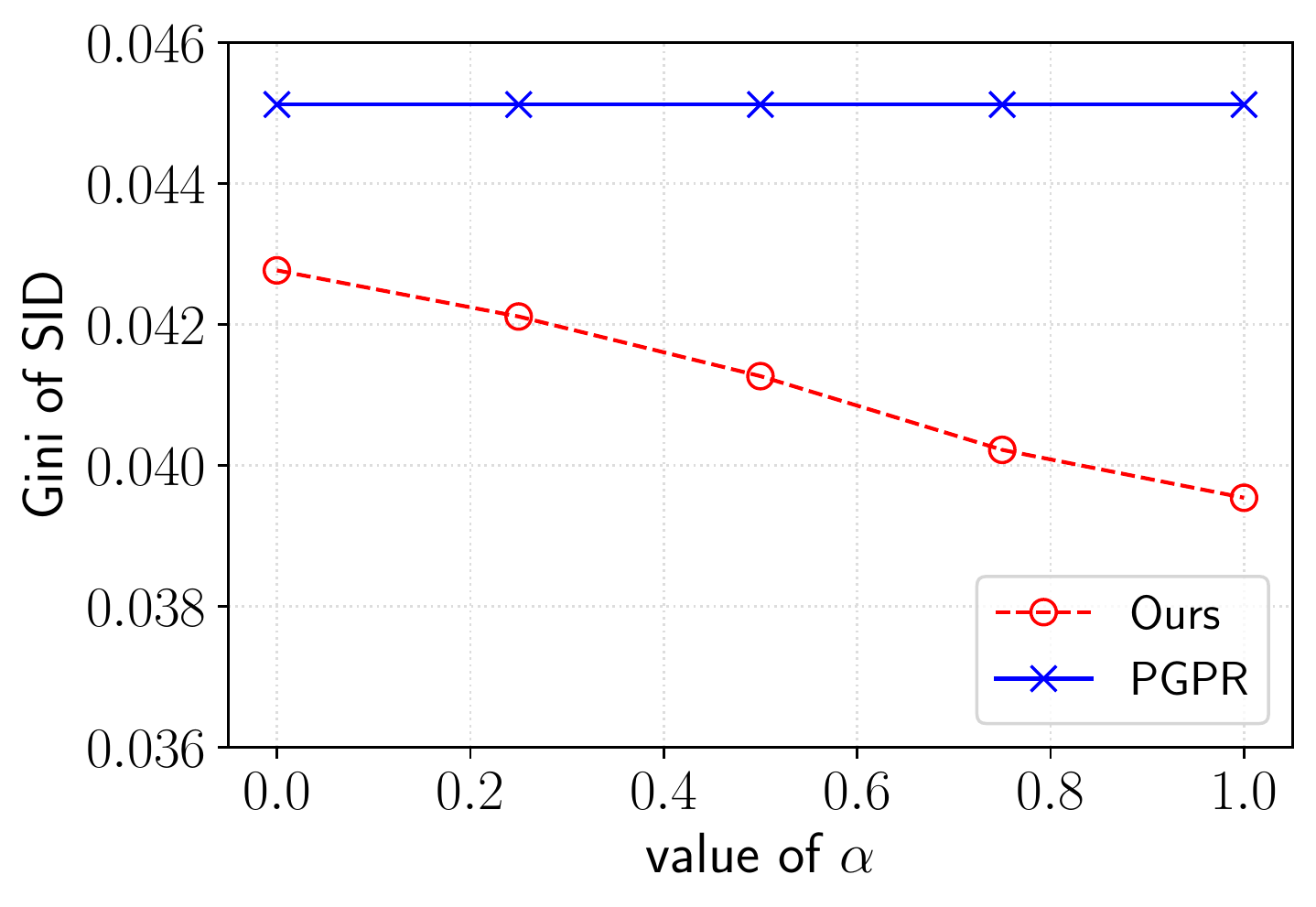}
\includegraphics[width=1.09in,height=0.85in]{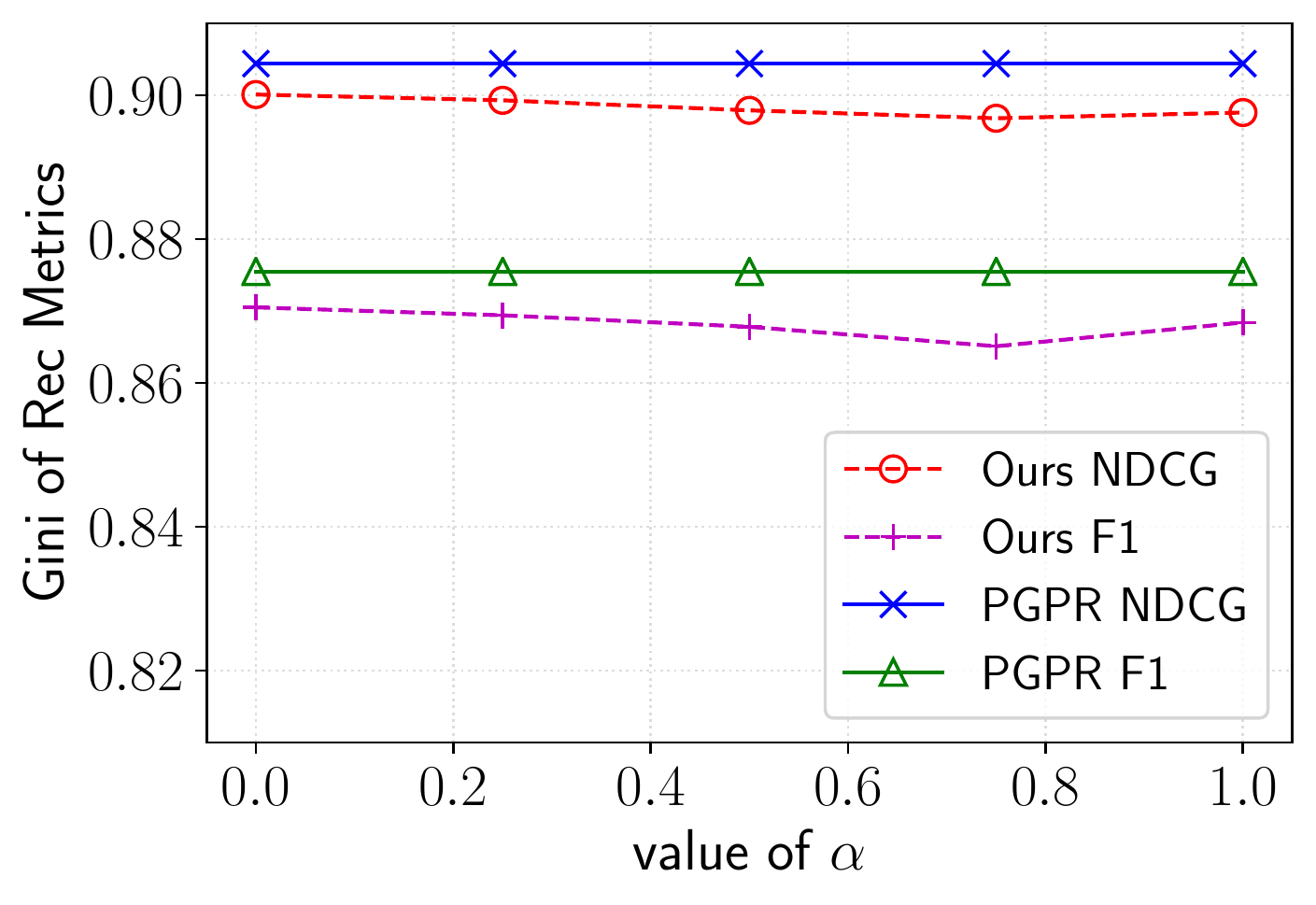}
\includegraphics[width=1.09in,height=0.85in]{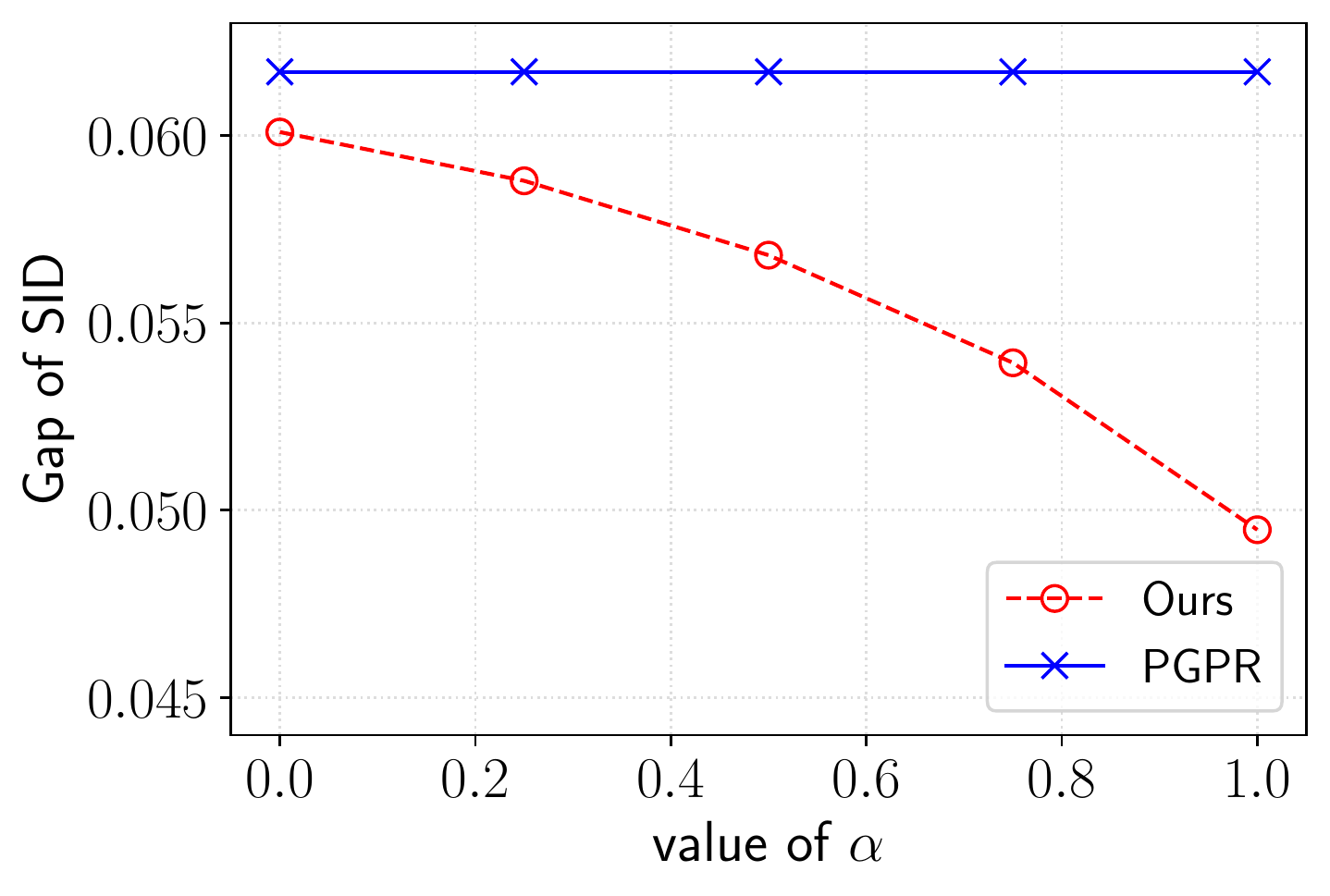} \\
\hspace*{0.2in} (a) Gini of $SID$ \hspace{0.15in} (b) Gini of Rec. Metrics  \hspace{0.1in} (c) Gap of $SID$
\vspace{3pt}
\includegraphics[width=1.09in,height=0.85in]{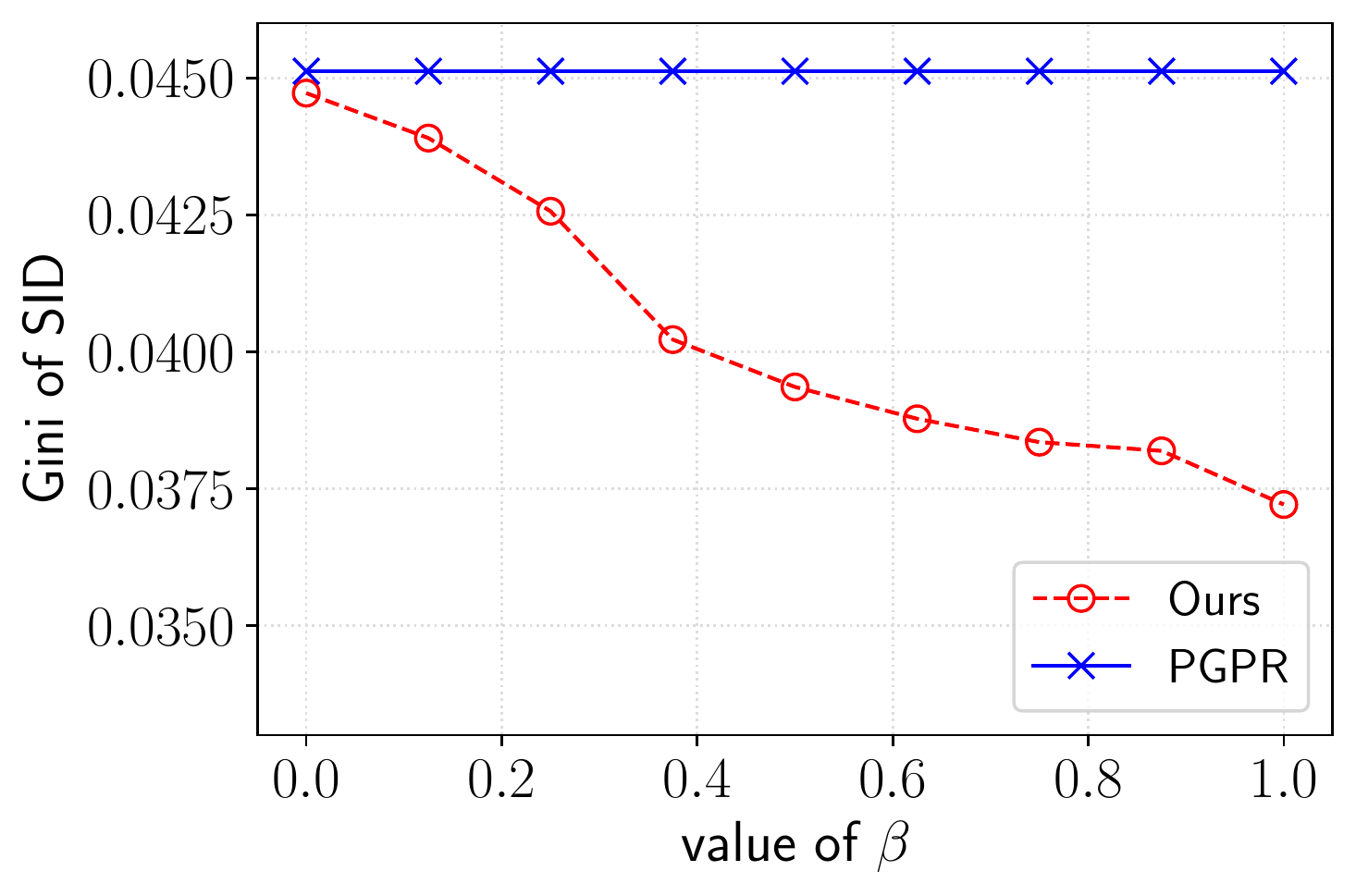}
\includegraphics[width=1.09in,height=0.85in]{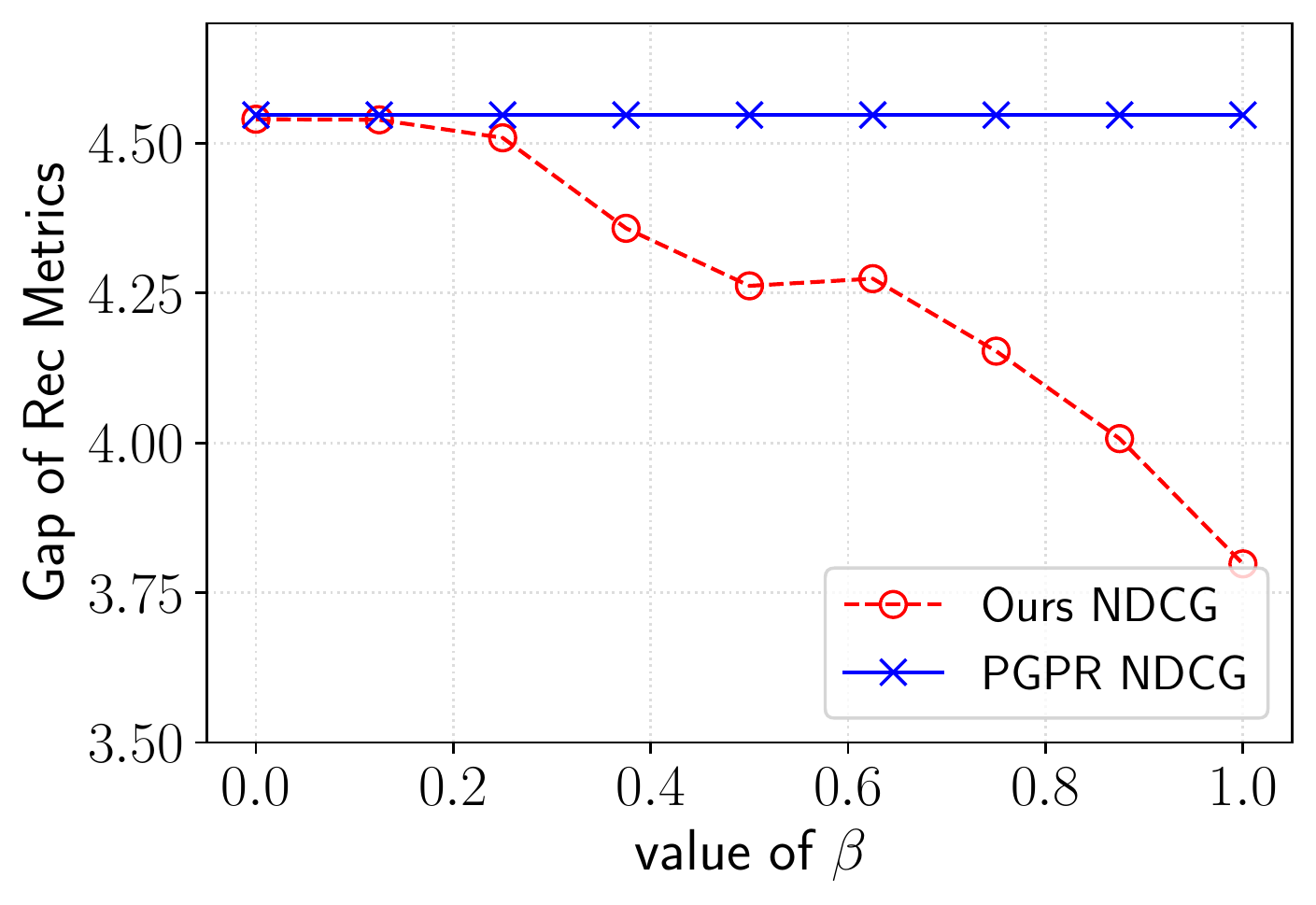}
\includegraphics[width=1.09in,height=0.85in]{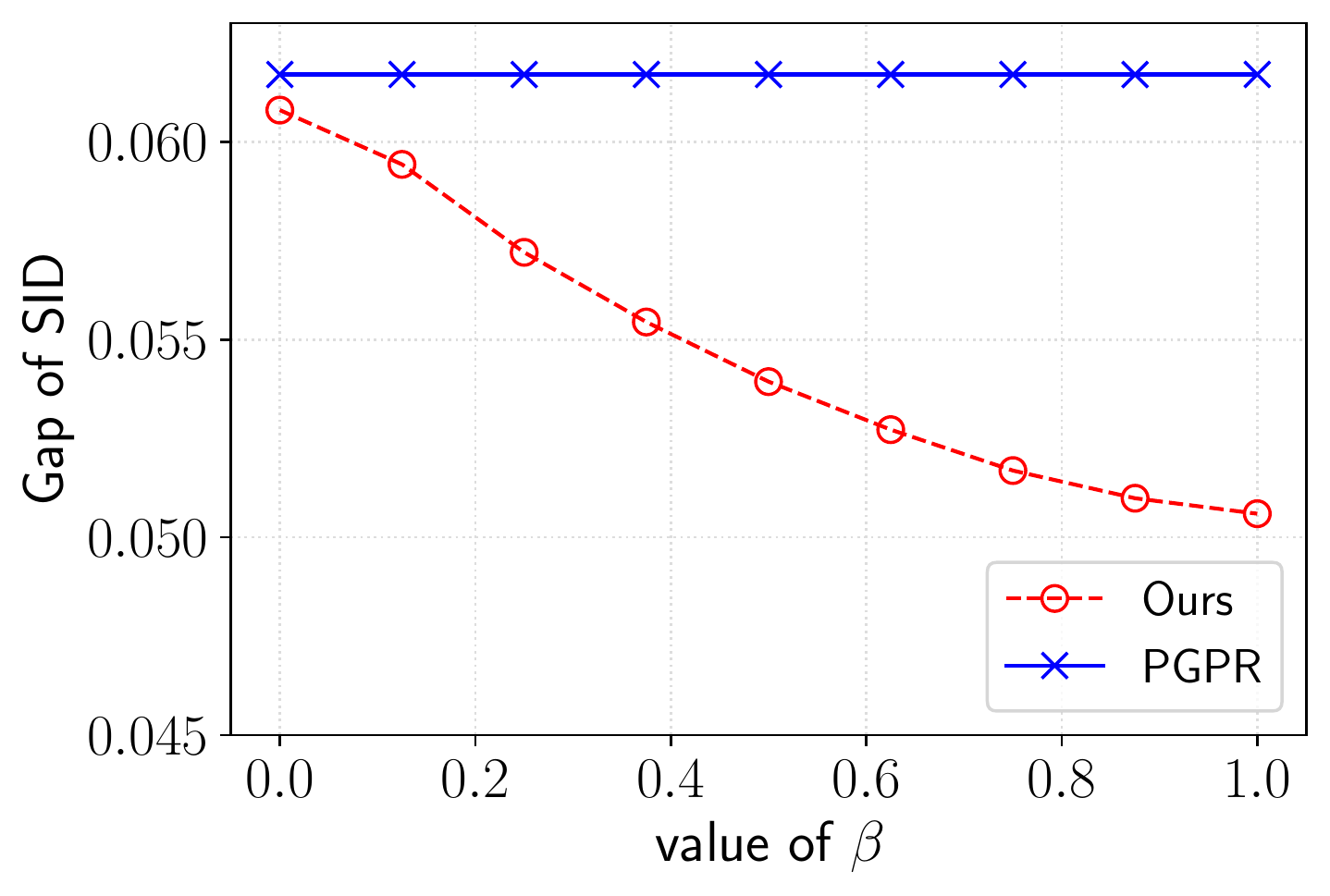} \\
\hspace*{-0.05in} (d) Gini of $SID$ \hspace{0.15in} (e) Gap of Rec. Metrics  \hspace{0.4in} (f) Gap of $SID$
\vspace{-5pt}
\caption{Results of Fairness-aware PGPR on Beauty dataset when fix $\beta=0.5$ (a-c) and fix $\alpha=0.75$ (d-f).}
\label{fig:PGPR_beauty}
\vspace{-7pt}
\end{figure}
\begin{figure}[t]
\centering
\includegraphics[width=1.09in,height=0.85in]{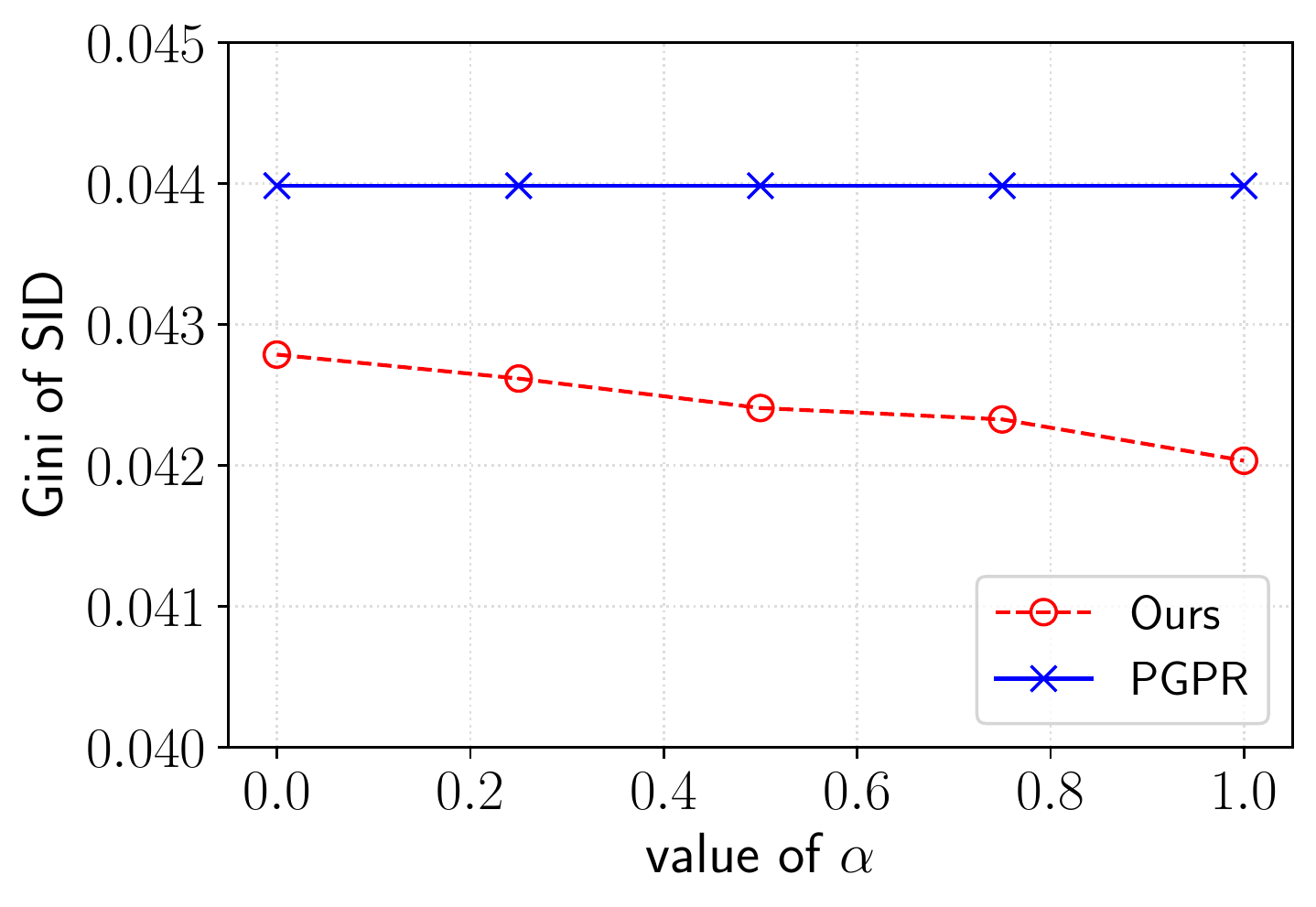}
\includegraphics[width=1.09in,height=0.85in]{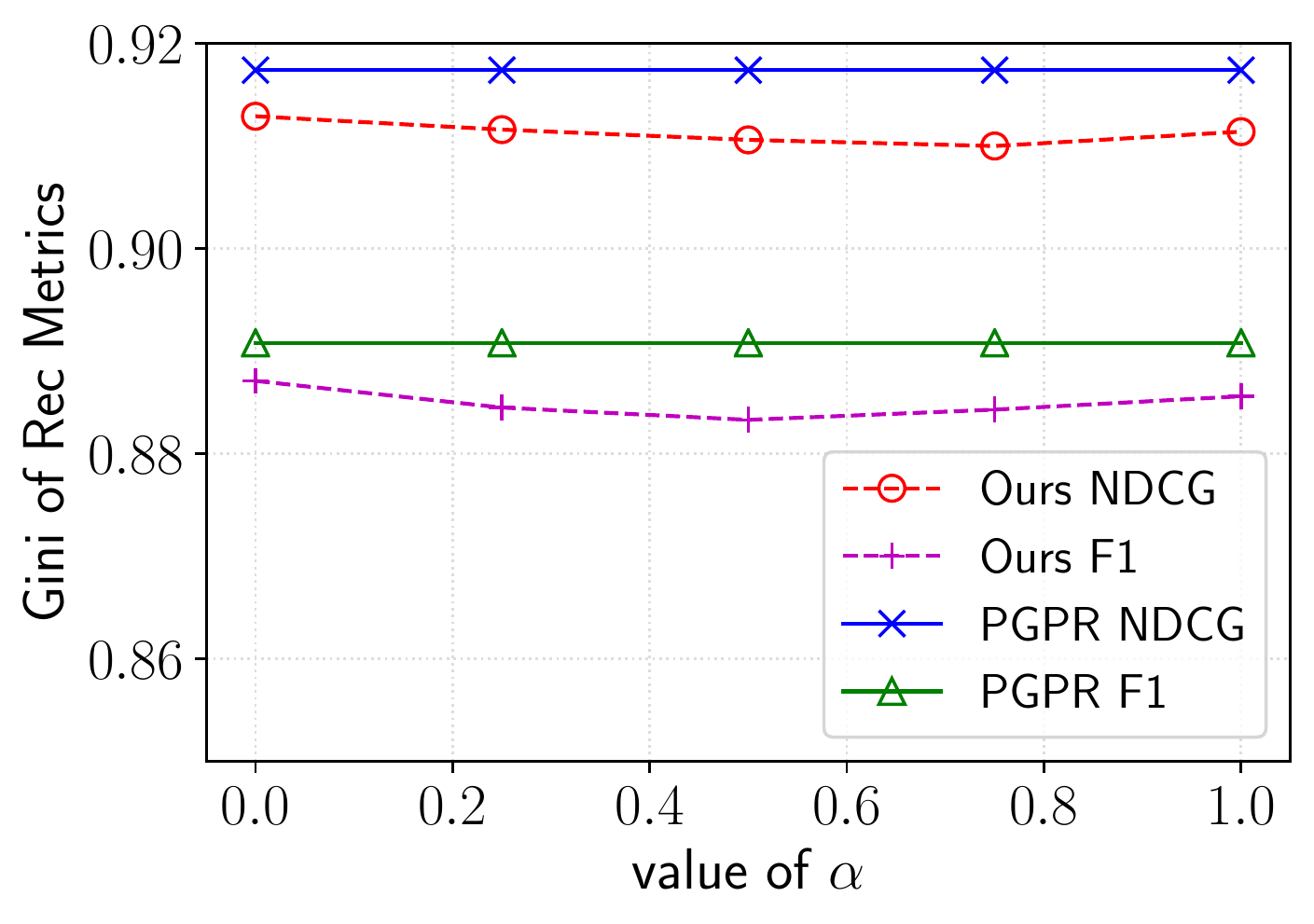}
\includegraphics[width=1.09in,height=0.85in]{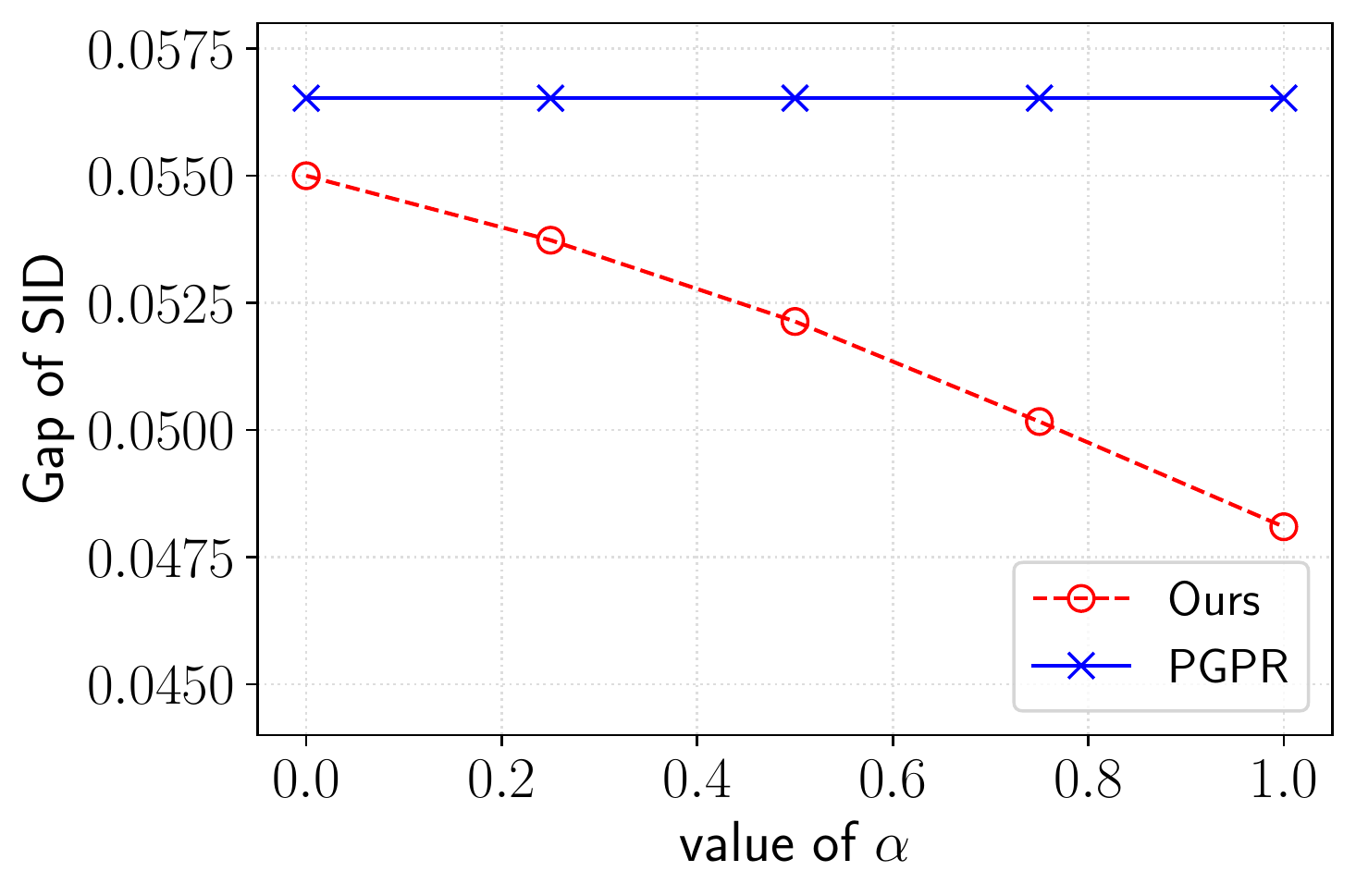} \\
\hspace*{0.2in} (a) Gini of $SID$ \hspace{0.15in} (b) Gini of Rec. Metrics  \hspace{0.1in} (c) Gap of $SID$
\vspace{3pt}
\includegraphics[width=1.09in,height=0.85in]{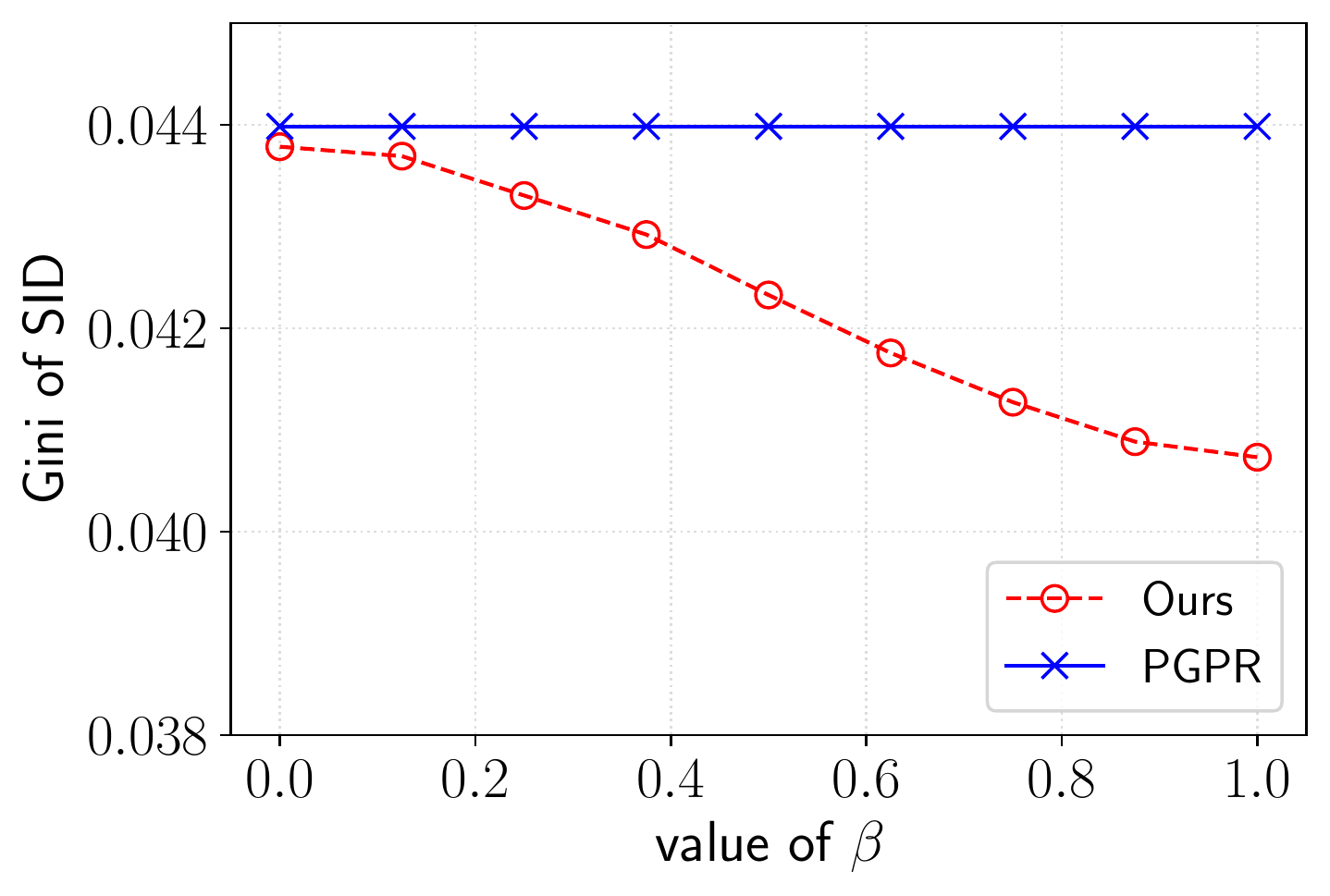}
\includegraphics[width=1.09in,height=0.85in]{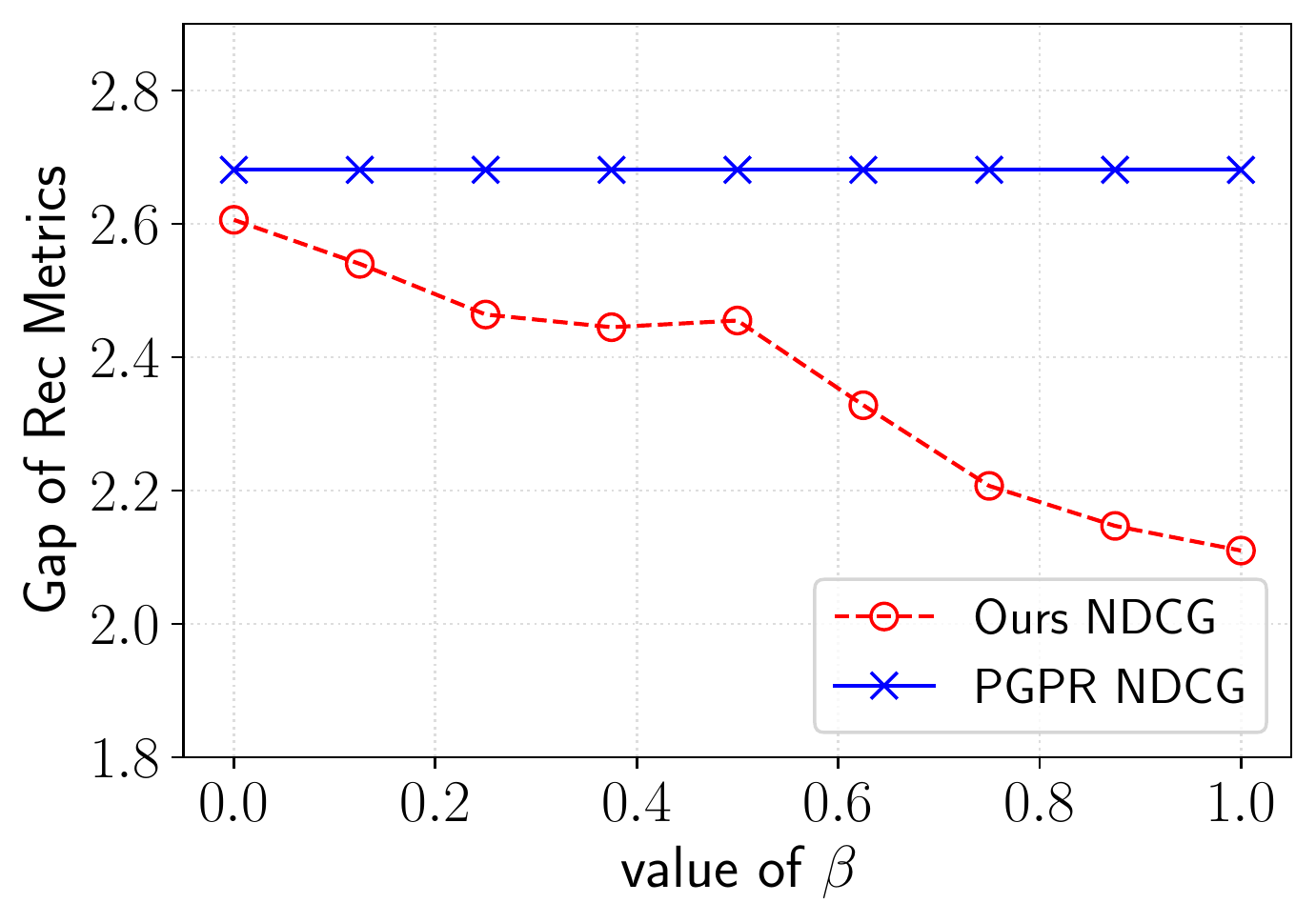}
\includegraphics[width=1.09in,height=0.85in]{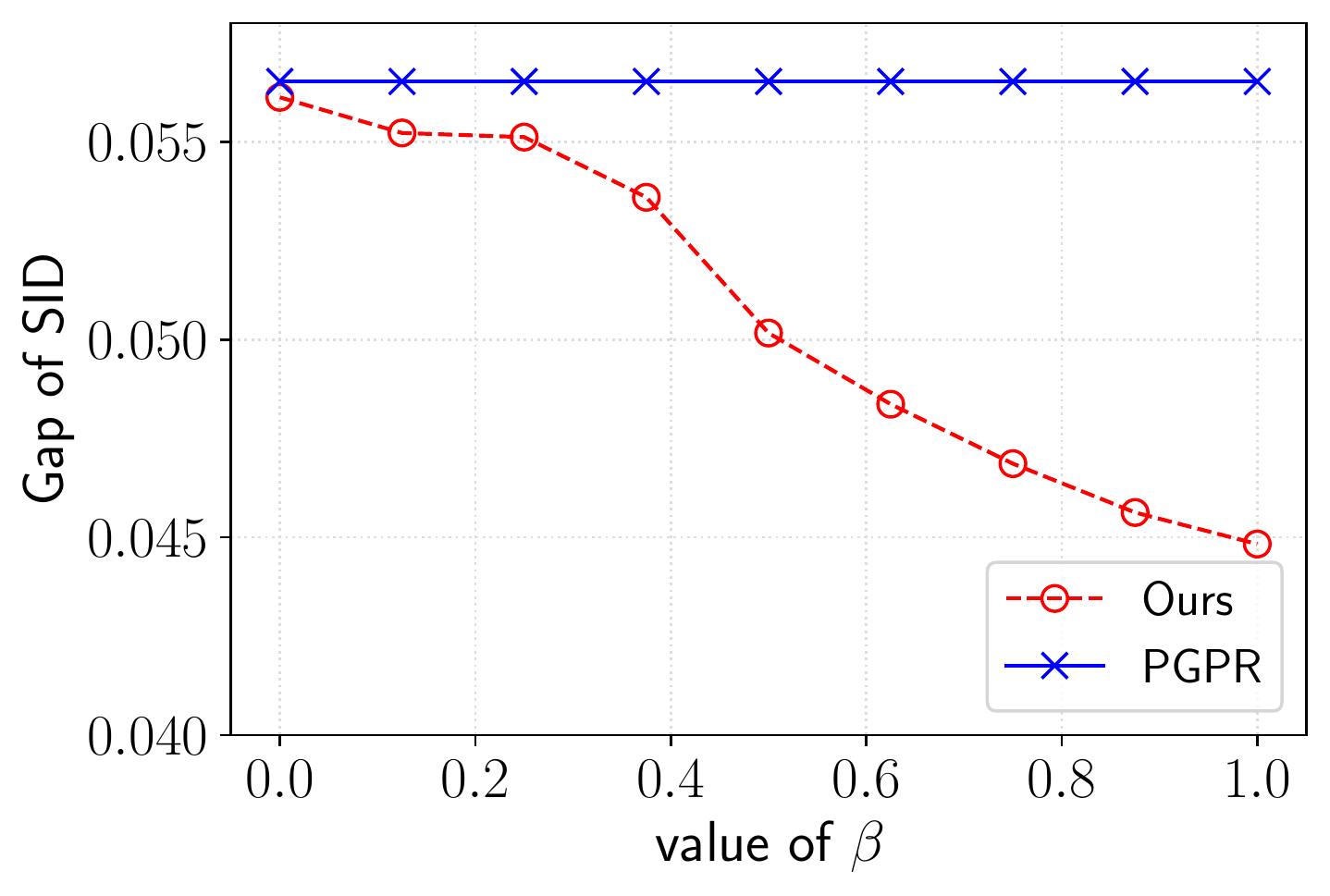} \\
\hspace*{-0.05in} (d) Gini of $SID$ \hspace{0.15in} (e) Gap of Rec. Metrics  \hspace{0.4in} (f) Gap of $SID$
\vspace{-5pt}
\caption{Results of Fairness-aware PGPR on Cellphone dataset when fix $\beta=0.5$ (a-c) and fix $\alpha=0.75$ (d-f).}
\label{fig:PGPR_cell}
\vspace{-5pt}
\end{figure}
\subsection{Main Results}
First, we show both the recommendation improvements and fairness effectiveness of our fairness-aware algorithm compared to state-of-the-art explainable RS models over KGs in terms of NDCG and F$_1$ scores as well as Group Recommendation Unfairness ($GRU$) between active group users and inactive group users.

The main results on the four Amazon datasets are provided in Table \ref{tab:eval}. Note that the \textbf{Difference} is calculated by the $GRU$ defined in Eq.~\ref{eq:GRU} through abstraction of corresponding metric scores between inactive users and active users. The overall performance is calculated by the 95\% of inactive users scores with the addition of 5\% of active users scores, which matches the ratio of the group split.\footnote{We report the results of the baselines with our own implementation. There is a slight discrepancy compared to original results.}
Our method outperforms all the baseline approaches for recommendation by a large margin in all settings. All of our fairness-aware algorithms take  uniform parameters with $\alpha = 0.75$ and $\beta = 0.5$. The overall performance of both approaches with fairness constraints even got boosted, while shrinking the $GRU$ fairness disparity between the groups. 
For example, on the Clothing dataset, our fairness-aware algorithm over PGPR achieves 3.101\% in NDCG, which is higher than 2.856\% of vanilla PGPR, and the group unfairness decreases to a great extent, from 1.410\% to 0.233\%. Similar trends can be observed for the other datasets.
Although we sacrifice some of the performance for the most active users, we substantially boost the performance for the inactive users. 
The disparity between inactive and active users gets closed for group fairness.

It is fairly remarkable
that after adopting the fairness-aware algorithm over two recent state-of-the-art baseline methods, we are able to retrieve substantially better recommendation results than the original methods.
We conjecture that our algorithm better harnesses the potential of current explainable recommendation methods by adapting the path distribution so that more diverse reasoning paths can be served to the users.
Since the inactive users undergo a prejudiced treatment due to the homogeneous user interactions with lower user visibility, they tend to benefit more under our fairness-aware algorithm.

\subsection{Ablation Study}
Besides recommendation performance, we also study how different weights for the fairness scores influence the fairness results regarding to the diversity metrics in Def.~\ref{def:GEDU} and the individual fairness metrics using the Gini coefficient defined in Defs.~\ref{def:IRU} and \ref{def:IEDU}.

\subsubsection{Study of fairness weight $\alpha$}
We first show how the choice of personalization weight $\alpha$ from  Eq.~\ref{eq:path_fairess_score} affects the recommendation performance, group fairness, and individual fairness. The results are plotted as (a-c) in Figs.~\ref{fig:HeteroEmbed_beauty}, \ref{fig:HeteroEmbed_cell}, \ref{fig:PGPR_beauty}, and \ref{fig:PGPR_cell}, including both our fairness-aware performance (red and purple curves) and the baselines (blue and green curves).

We observe that our model consistently outperforms vanilla HeteroEmbed and PGPR baselines under all settings of $\alpha$ in terms of Group Explainable Diversity Unfairness ($GEDU$) and individual Recommendation Unfairness ($IRU$), as shown in parts (a) and (c) of the aforementioned figures. The unfairness is minimized at the point of $\alpha=1.0$, i.e., when not accounting for historical path distributions, but focusing solely on debiasing the intrinsic user preferences, our model can achieve a high degree of fairness.
However, with an apt choice of $\alpha$, we can not only benefit the $IRU$ in (b) figures, but also maintain the more reliable $GEDU$ and $IRU$.   

\subsubsection{Study of ranking weight $\beta$}
Next, we consider the effect of the ranking weight $\beta$ from  Eq.~\ref{eq:rank_fairess_socre}. 
Since we already obtain the $K$ items from the filtered original top-$N$ list, the $\beta$ factor does not change the F$_1$ score. We provide results on the Cellphones and Beauty datasets, first fixing $\alpha=0.75$, which was shown to give strong results in the main results.

Subfigures (d-f) in Figs.~\ref{fig:HeteroEmbed_beauty}, \ref{fig:HeteroEmbed_cell},  \ref{fig:PGPR_beauty}, and \ref{fig:PGPR_cell} plot the results. We observe, first of all, that as the weight of $\beta$ increases, our fairness-aware algorithm is able to consistently reduce the $GEDU$ and $IEDU$. For $\beta = 0$, there is no external fairness weight incorporated into the overall ranking score, which obtains slightly better results than the baselines owing to the optimization process.
In contrast, $\beta = 1.0$ represents only considering the fairness weights but ignoring the original recommendation scores.
The optimal choice for minimizing the $IRU$ is when $\beta$ is around 0.2 for Fair HeteroEmbed, while for Fair PGPR, $\beta=1$ obtains better results. This might be because the PGPR method prunes away some less salient word features via TF-IDF scoring.
In such cases, paths containing the feature word will be eliminated. Also, some inactive users prefer making comments on items rather than purchasing them. Such preprocessing will make the inactive users hold even less interactions. Therefore, our fairness-aware explainable algorithm yields strong fairness improvements when $\beta=1$ for the PGPR baseline.
\vspace{-10pt}
\subsection{Study of Recommendation Quality under Fairness-Aware Algorithm}
After studying the fairness metrics, we next study how the recommendation quality is affected by the parameter choices.
From Eqs.~\ref{eq:path_fairess_score} and \ref{eq:rank_fairess_socre}, we can infer that changes of $\beta$ can exert more influence than $\alpha$. As we fix the path fairness weight $\alpha=0.75$, the right side of Table \ref{tab:path_fariness_exp} indicates the recommendation performance of inactive users initially is boosted as $\beta$ grows, and then the performance starts to drop as $\beta$ approaches $1$. 
Similar conclusions can be drawn: If there is less weight on the recommendation score $\mathbf{\hat Q_{i}}$ as a guide to create a proper ranking of recommended lists and balance the fairness effects, the role of generated explainable paths would gradually diminish, leading the model towards making inappropriate ranking decisions.

The left part of Table~\ref{tab:path_fariness_exp} reflects the recommendation performance variance in terms of path fairness weight $\alpha$. As we discussed before, $\alpha=0$ suggests the fairness-aware algorithm is no longer taking advantage of the user debiasing. As the ratio of $\alpha$ increases, the larger the effect of the self adjustment regularization, which leads to a decreasing $GRU$. However, when $\alpha=1$, both the recommendation performance and the $GRU$ decrease at a small rate. We can conclude that the zero score of path diversity mean that the weight of the generated path diversity has been eliminated, which leads to fairness scores only coming from the path ranking component. Thus, the recommendation performance degrades to a small extent.
The results appear reasonable because the model naturally will behave  less intelligently if it either completely ignores the user's historical interactions or if it considers only the path 
debiasing. Any extreme regularization of one particular side may harm the recommendation performance. Therefore, it is indispensable to choose the proper hyper-parameters for robust performance.

Still, note that no matter how we change $\alpha$ or $\beta$, in Table \ref{tab:path_fariness_exp}, we can observe that the group unfairness metric $GRU$ is always better than for the baselines.

\begin{table}[t]
\renewcommand\arraystretch{1.1}
\setlength{\tabcolsep}{3pt}
\resizebox{1.0\linewidth}{!}{
\begin{tabular}{l|cccccc|ccc}
\cline{1-10}
\cline{1-10}
\cline{1-10}
\cline{1-10}
\multicolumn{1}{c|}{} & \multicolumn{6}{c|}{\textbf{Beauty}} & \multicolumn{3}{c}{\textbf{Cell Phones}}\\
\cline{1-10}
\multirow{2}{*}{Metric($\%$)}
 & \multicolumn{2}{c}{Inactive (IA)} & \multicolumn{2}{c}{Active (A)} & \multicolumn{2}{c|}{GRU} & IA & A & GRU\\
 & {\small NDCG}  & F$_1$ & {\small NDCG}    & F$_1$  & {\small NDCG}  & F$_1$ & {\small NDCG}  & {\small NDCG}  & {\small NDCG}  \\
\cline{1-10}
(*)     & 6.075 & 2.756 & 11.930 & 10.132& 5.855 & 7.376 & 5.645 & 9.395 & 3.75  \\
0.0     & 6.317 & 2.916 & 12.088 & 9.765 & 5.771 & 6.949 & 5.658 & 8.821 & 3.163 \\
0.125   & 6.354 & 2.926 & 12.088 & 9.790 & 5.734 & 6.864 & 5.782 & 8.888 & 3.106 \\
0.25    & 6.407 & 2.930 & 12.089 & 9.800 & 5.682 & 6.870 & 5.924 & 8.999 & 3.075 \\
0.375   & 6.459 & 2.933 & 12.206 & 9.802 & 5.747 & 6.869 & 6.033 & 9.254 & 3.221 \\
0.5     & 6.470 & 2.935 & 12.232 & 9.808 & 5.762 & 6.873 & 6.037 & 9.284 & 3.247 \\
0.625   & 6.462 & 2.928 & 12.230 & 9.834 & 5.768 & 6.906 & 5.974 & 9.216 & 3.242 \\
0.75    & 6.451 & 2.924 & 12.229 & 9.853 & 5.778 & 6.929 & 5.782 & 9.093 & 3.211 \\
0.875   & 6.398 & 2.872 & 12.191 & 9.776 & 5.793 & 6.904 & 5.479 & 8.667 & 3.188 \\
1.0     & 6.368 & 2.836 & 12.145 & 9.732 & 5.777 & 6.891 & 5.217 & 8.373 & 3.156 \\
\cline{1-10}
\cline{1-10}
\cline{1-10}
\cline{1-10}
\end{tabular}
}
\caption{Left: Performance under different ratios of $\alpha$ for Fairness-aware HeteroEmbed with fixed $\beta=0.5$. Right: Performance under different ratios of $\beta$ for Fairness-aware HeteroEmbed with fixed $\alpha=0.75$. (*) is baseline performance
}
\label{tab:path_fariness_exp}
\vspace{-20pt}
\end{table}

\subsection{Case study}
\begin{figure}
\centering
\includegraphics[width=0.99\linewidth]{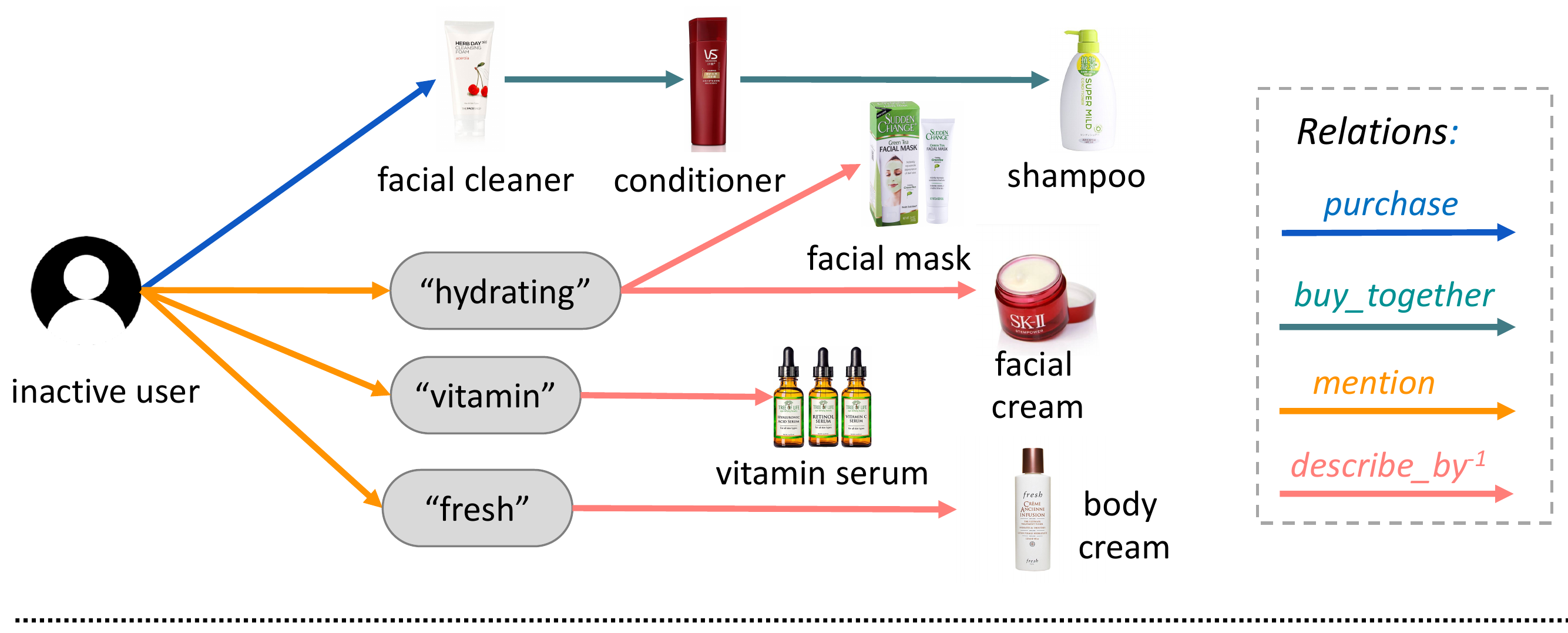}\\
\includegraphics[width=0.99\linewidth]{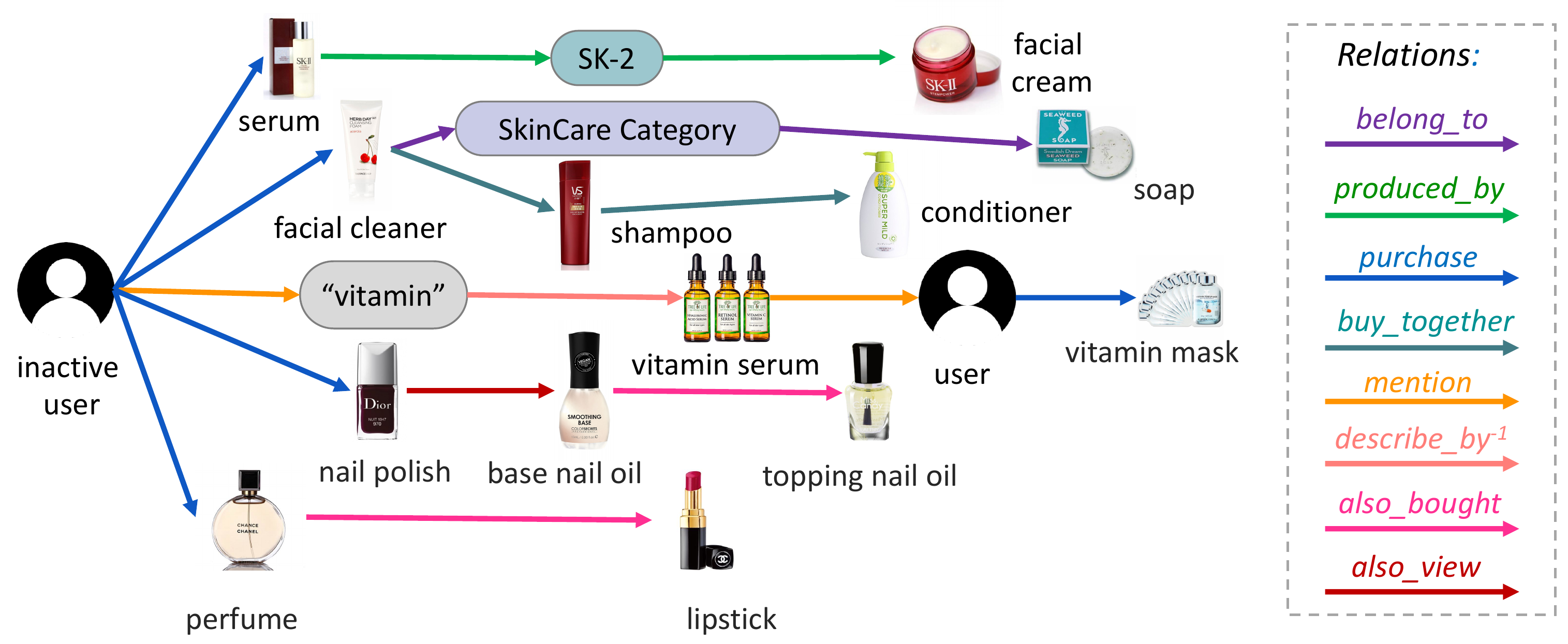}
\caption{Case study of real recommendation paths, before and after adding our fairness algorithm.}
\label{fig:cases}
\vspace{-10pt}
\end{figure}

Figure \ref{fig:cases} illustrates the real effects of our fairness-aware algorithm in terms of path diversity and the accuracy of predicted items. The example comes from the Beauty dataset. The upper part illustrates the outputs of the original vanilla PGPR explainable RS method, which neglects path diversity. It is filled with ``user--mention'' paths connecting to the predicted items with only 4 items appearing in the top-$k$ list. For comparison, after adding our fairness-aware algorithm, the model predicts 6 correct items that users are willing to purchase, associating them with diverse explainable paths. Hence, our approach is able to invoke more items, finding alternative kinds of user--item paths. At the same time, considering the ``facial cleaner'' example, our fairness-aware algorithm considers two further items connecting with it, so that the user is able to buy further related items. This shows how our fairness-aware method not only is capable of considering a more diverse and comprehensive set of explainable paths, but also ends up finding more correct recommendations.

\section{Conclusions}
In this work, we study the prominent problem of fairness in the context of state-of-the-art explainable recommendation algorithms over knowledge graphs. We first quantify unfairness at both the individual level and the group level. Based on this, we propose fairness metrics in terms of path diversity as well as the recommendation performance disparity. We then present a generalized fairness-aware algorithm that is capable not only of reducing the disparity but also of maintaining the recommendation quality. We extensively evaluate our model on several real-world datasets, and demonstrate that our approach reduces unfairness by providing diverse path patterns and strong explainable recommendation results.
The source code of our work at \url{https://github.com/zuohuif/FairKG4Rec} is publicly available.

\section{Acknowledgement}
The authors thank Dr. Longqi Yang for many enlightening conversations and the reviewers for the valuable comments and constructive suggestions. This work was supported in part by NSF IIS-1910154. Any opinions, findings, conclusions or recommendations expressed in this material are those of the authors and do not necessarily reflect those of the sponsors.

\bibliographystyle{ACM-Reference-Format}
\balance
\bibliography{paper}

\end{document}